\documentclass[useAMS,usenatbib]{mn2e} 
\usepackage{graphicx}
\usepackage{amsmath}
\usepackage{amssymb}
\usepackage[font={it},labelfont={bf}]{caption}
\usepackage{units}
\usepackage{times}
\usepackage{color}
\usepackage[usenames,dvipsnames,svgnames,table]{xcolor}

\usepackage{fixltx2e} 

\definecolor{customhdrcolor}{rgb}{0.0,0.0,0.0}
\definecolor{customcitecolor}{rgb}{0.0,0.5,0.75}
\definecolor{customlinkcolor}{rgb}{0.0,0.5,0.75}

\usepackage[colorlinks=true,linkcolor=customlinkcolor,urlcolor=customlinkcolor,citecolor=customcitecolor,pdftex]{hyperref}

\ifpdf\else\usepackage{graphics}\fi

\DeclareRobustCommand{\TUSSEN}[3]{#2}

\title[Brightness and spatial distributions of terrestrial radio sources]{The brightness and spatial distributions of terrestrial radio sources}

\author[A.~R.~Offringa et al.]{A.~R.~Offringa$^{1,2,3}$\thanks{E-mail:
\url{offringa@mso.anu.edu.au}},
A.~G.~de~Bruyn$^{4,3}$, S.~Zaroubi$^{3}$, L.~V.~E.~Koopmans$^{3}$, \newauthor
S.~J.~Wijnholds$^{4}$, F.~B.~Abdalla$^{5}$, W.~N.~Brouw$^{3,4}$, B.~Ciardi$^{6}$, I.~T.~Iliev$^{7}$, \newauthor
G.~J.~A.~Harker$^{8}$, G.~Mellema$^{9}$, G.~Bernardi$^{10}$, P.~Zarka$^{11}$, A.~Ghosh$^{3}$,\newauthor A.~Alexov$^{12}$, J.~Anderson$^{13}$, A.~Asgekar$^{4}$, I.~M.~Avruch$^{14,3}$, R.~Beck$^{13}$,\newauthor
M.~E.~Bell$^{2,15}$, M.~R.~Bell$^{6}$, M.~J.~Bentum$^{4}$, P.~Best$^{16}$, L.~B\^{i}rzan$^{17}$, F.~Breitling$^{18}$,\newauthor
J.~Broderick$^{19}$, M.~Br\"uggen$^{20}$, H.~R.~Butcher$^{4,1}$, F.~de~Gasperin$^{20}$, E.~de~Geus$^{4}$,\newauthor
M.~de~Vos$^{4}$, S.~Duscha$^{4}$, J.~Eisl\"offel$^{21}$, R.~A.~Fallows$^{4}$, C.~Ferrari$^{22}$,\newauthor
W.~Frieswijk$^{4}$, M.~A.~Garrett$^{4,17}$, J.~Grie\ss{}meier$^{23}$, T.~E.~Hassall$^{19}$, A.~Horneffer$^{13}$,\newauthor
M.~Iacobelli$^{17}$, E.~Juette$^{24}$, A. ~Karastergiou$^{25}$, W.~Klijn$^{4}$, V.~I.~Kondratiev$^{4,26}$,\newauthor
M.~Kuniyoshi$^{13}$, G.~Kuper$^{4}$, J.~van~Leeuwen$^{4,27}$, M.~Loose$^{4}$, P.~Maat$^{4}$,\newauthor
G.~Macario$^{22}$, G.~Mann$^{18}$, J.~P.~McKean$^{4}$, H.~Meulman$^{4}$, M.~J.~Norden$^{4}$,\newauthor
E.~Orru$^{4}$, H.~Paas$^{28}$, M.~Pandey-Pommier$^{29}$, R.~Pizzo$^{4}$, A.~G.~Polatidis$^{4}$,\newauthor
D.~Rafferty$^{17}$, W.~Reich$^{13}$, R.~van~Nieuwpoort$^{4}$, H.~R\"ottgering$^{17}$,\newauthor
A.~M.~M.~Scaife$^{19}$, J.~Sluman$^{4}$, O.~Smirnov$^{30,31}$, C.~Sobey$^{13}$, M.~Tagger$^{23}$,\newauthor
Y.~Tang$^{4}$, C.~Tasse$^{11}$, S.~ter~Veen$^{32}$, C.~Toribio$^{4}$, R.~Vermeulen$^{4}$, C.~Vocks$^{18}$,\newauthor
R.~J.~van~Weeren$^{10}$, M.~W.~Wise$^{4,27}$, O.~Wucknitz$^{33,13}$
\\
$^{1}$RSAA, Australian National University, Mt Stromlo Observatory, via Cotter Road, Weston, ACT 2611, Australia \\
$^{2}$ARC Centre of Excellence for All-sky Astrophysics (CAASTRO) \\
$^{3}$Kapteyn Astronomical Institute, PO Box 800, 9700 AV Groningen, The Netherlands \\
$^{4}$Netherlands Institute for Radio Astronomy (ASTRON), Postbus 2, 7990 AA Dwingeloo, The Netherlands \\
$^{5}$UCL Department of Physics and Astronomy, London WC1E 6BT, United Kingdom \\
$^{6}$Max Planck Institute for Astrophysics, Karl Schwarzschild Str. 1, 85741 Garching, Germany \\
$^{7}$University of Sussex, Falmer, Brighton BN1 9QH, UK \\
$^{8}$Center for Astrophysics and Space Astronomy, University of Colorado Boulder, CO 80309, USA \\
$^{9}$Stockholm University, AlbaNova University Center, Stockholm Observatory, SE-106 91 Stockholm, Sweden \\
$^{10}$Harvard-Smithsonian Center for Astrophysics, 60 Garden Street, Cambridge, MA 02138, USA \\
$^{11}$LESIA, UMR CNRS 8109, Observatoire de Paris, 92195   Meudon, France \\
$^{12}$Space Telescope Science Institute, 3700 San Martin Drive, Baltimore, MD 21218, USA \\
$^{13}$Max-Planck-Institut f\"{u}r Radioastronomie, Auf dem H\"ugel 69, 53121 Bonn, Germany \\
$^{14}$SRON Netherlands Insitute for Space Research, Sorbonnelaan 2, 3584 CA, Utrecht, The Netherlands \\
$^{15}$Sydney Institute for Astronomy, School of Physics, The University of Sydney, NSW 2006, Australia \\
$^{16}$Institute for Astronomy, University of Edinburgh, Royal Observatory of Edinburgh, Blackford Hill, Edinburgh EH9 3HJ, UK \\
$^{17}$Leiden Observatory, Leiden University, PO Box 9513, 2300 RA Leiden, The Netherlands \\
$^{18}$Leibniz-Institut f\"{u}r Astrophysik Potsdam (AIP), An der Sternwarte 16, 14482 Potsdam, Germany \\
$^{19}$School of Physics and Astronomy, University of Southampton, Southampton, SO17 1BJ, UK \\
$^{20}$University of Hamburg, Gojenbergsweg 112, 21029 Hamburg, Germany \\
$^{21}$Th\"{u}ringer Landessternwarte, Sternwarte 5, D-07778 Tautenburg, Germany \\
$^{22}$Laboratoire Lagrange, UMR7293, Universit\`{e} de Nice Sophia-Antipolis, CNRS, Observatoire de la C\'{o}te d'Azur, 06300 Nice, France \\
$^{23}$Laboratoire de Physique et Chimie de l' Environnement et de l' Espace, LPC2E UMR 7328 CNRS, 45071 Orl\'{e}ans Cedex 02, France \\
$^{24}$Astronomisches Institut der Ruhr-Universit\"{a}t Bochum, Universitaetsstrasse 150, 44780 Bochum, Germany \\
$^{25}$Astrophysics, University of Oxford, Denys Wilkinson Building, Keble Road, Oxford OX1 3RH \\
$^{26}$Astro Space Center of the Lebedev Physical Institute, Profsoyuznaya str. 84/32, Moscow 117997, Russia \\
$^{27}$Astronomical Institute 'Anton Pannekoek', University of Amsterdam, Postbus 94249, 1090 GE Amsterdam, The Netherlands \\
$^{28}$Center for Information Technology (CIT), University of Groningen, The Netherlands \\
$^{29}$Centre de Recherche Astrophysique de Lyon, Observatoire de Lyon, 9 av Charles Andr\'{e}, 69561 Saint Genis Laval Cedex, France \\
$^{30}$Centre for Radio Astronomy Techniques \& Technologies (RATT), Department of Physics and Elelctronics, Rhodes University, PO Box 94, Grahamstown 6140, South Africa \\
$^{31}$SKA South Africa, 3rd Floor, The Park, Park Road, Pinelands, 7405, South Africa \\
$^{32}$Department of Astrophysics/IMAPP, Radboud University Nijmegen, P.O. Box 9010, 6500 GL Nijmegen, The Netherlands \\
$^{33}$Argelander-Institut f\"{u}r Astronomie, University of Bonn, Auf dem H\"{u}gel 71, 53121, Bonn, Germany\vspace*{-0.5cm}
}

\begin{document}

\date{Accepted 2013 July 16. Received 2013 July 16; in original form 2013 March 1}
\pagerange{\pageref{firstpage}--\pageref{lastpage}}
\pubyear{2013}

\label{firstpage}
\maketitle

\begin{abstract}
Faint undetected sources of radio-frequency interference (RFI) might become visible in long radio observations when they are consistently present over time. Thereby, they might obstruct the detection of the weak astronomical signals of interest. This issue is especially important for Epoch of Reionisation (EoR) projects that try to detect the faint redshifted HI signals from the time of the earliest structures in the Universe. We explore the RFI situation at 30--163~MHz by studying brightness histograms of visibility data observed with LOFAR, similar to radio-source-count analyses that are used in cosmology. An empirical RFI distribution model is derived that allows the simulation of RFI in radio observations. The brightness histograms show an RFI distribution that follows a power-law distribution with an estimated exponent around -1.5. With several assumptions, this can be explained with a uniform distribution of terrestrial radio sources whose radiation follows existing propagation models. Extrapolation of the power law implies that the current LOFAR EoR observations should be severely RFI limited if the strength of RFI sources remains strong after time integration. This is in contrast with actual observations, which almost reach the thermal noise and are thought not to be limited by RFI. Therefore, we conclude that it is unlikely that there are undetected RFI sources that will become visible in long observations. Consequently, there is no indication that RFI will prevent an EoR detection with LOFAR.
\end{abstract}


\section{Introduction}
Radio astronomy concerns itself with the observation of radiation from celestial sources at radio wavelengths. However, astronomical radio observations can be affected by radio-frequency interference (RFI), which might make it difficult to calibrate the instrument and achieve high sensitivities \citep{impact-of-warc79,interference-and-radioastronomy-1991,interference-model-lemmon,rfi-mitigation-overview-fridman-baan}. The careful management of spectrum allocation and the construction of radio-quiet zones help to limit the number of harmful transmitters. If harmful RFI is observed nevertheless, the use of RFI mitigation methods can sometimes clean the data sufficiently to allow succesful calibration and imaging. Many techniques have been designed to mitigate the effects of RFI, such as detection and flagging of data \citep{chi-square-time-blanking-weber, multichannel-rfi-mitigation, exoplanet-detection-with-rfi, wsrt-rfims, pulse-blanking, effelsberg-rfi-mitigation, post-correlation-rfi-classification}, adaptive cancellation techniques \citep{adaptive-cancellation,post-correlation-reference-signal} and spatial filtering \citep{multichannel-rfi-mitigation, ellingson-spatial-nulling-2002, hampson-spatial-nulling-2002, boonstra-dissertation, spatial-filtering-parkes-multibeam-for-pulses, post-correlation-filtering}. 

Typical radio observations record a few hours of data, and the results are integrated. In these cases, excising only the interference that is apparent and thus above the noise often suffices, i.e., the observation can still reach the thermal noise limit of the instrument. A new challenge arises, however, when one desires much deeper observations, and hundreds of hours of observations need to be integrated. In such a case, weak interference caused by stationary RFI sources might not manifest itself above the noise in individual observations, but might be persistently present in the data. Subsequently, when averaging these data, the interference might become apparent and occlude the signal of interest. This is very relevant for the 21-cm Epoch of Reionisation (EoR) experiments, because they involve long integration times. Several such experiments are underway, to either measure the angular power spectrum \citep{gmrt-eor-2011-paciga,de-bruyn-eor-ursi-2011,eor-paper-2011-jacobs,eor-mwa-2012-williams} or the global signal \citep{edges}. Ground-based Cosmic Microwave Background (CMB) experiments are another class of experiments involving long integration times (e.g., \citealt{ska-cmb-subrahmanyan}). For these experiments, it is important to know the possible effect of low-level interference on the data, as these might overshadow or alter the signal of interest.

In this article, we will connect new insights about RFI to the angular EoR experiment that is using the Low-Frequency Array (LOFAR) (\citealt{de-bruyn-eor-ursi-2011}, \citealt{lofar-overview-2013}). The LOFAR EoR project aims to detect the redshifted 21-cm signals from the EoR using the LOFAR HBA antennas (115--190~MHz, $z_\textrm{HI}$=11.4--6.5). Several fields will be observed over 100~nights, to achieve sufficient sensitivity to allow the signal extraction. An EoR calibration pipeline has been designed that solves for ionospheric and instrumental effects in approximately hundred directions using the SAGE algorithm \citep{sage-calibration-ii}. Initial results from commissioning observations show that in a single night the thermal noise level can almost be reached \citep{ncp-eor-yatawatta}.

This work explores the information that is present in interference distributions, in order to analyse possible low-level interference that is not detectable by standard detection methods. Our approach is similar to the radio-source-count analyses that are used in cosmology \citep{condon-cosm-evol-of-radio-sources}, also named $\log N$ -- $\log S$ analyses, where $N$ and $S$ refer to the celestial source count and brightness respectively. The slope in such a plot contains information about source populations, their luminosity functions and the geometry of the Universe. We analyse such a double-logarithmic plot for the case of terrestrial sources, with the ultimate goal of estimating their full spatial and brightness distributions. This results in a better insight into the effects of low-level interference and allows one to simulate the effects of interference more accurately.


This paper is organised as follows: in Sect.~\ref{sec:distribution-prediction}, we calculate a model for terrestrial interfering source distribution based on various assumptions. Sect.~\ref{sec:distribution-methods} presents the methods that we use to generate and analyse brightness histograms of LOFAR data. Sect.~\ref{sec:dist-data} describes the two LOFAR data sets that have been used to perform the experiment. The results of analysing the sets are presented in Sect.~\ref{sec:dist-results}. Finally, in Sect.~\ref{sec:dist-discussion} the results are discussed and conclusions are drawn.

\section{Modelling the brightness distribution} \label{sec:distribution-prediction}
Interference is generated by many different kinds of transmitters, and these will have different spatial and brightness distributions (``spatial'' refers here to the distribution on the Earth). For example, aeroplanes and satellites have widely different heights, while other sources are ground-based. Even ground-based sources might be spread differently. For example, it can be expected that citizens' band (CB) devices, that are often used in cars, are distributed differently from broadcasting transmitters. For deliberate transmitters, the frequency at which interference occurs can identify the involved class of devices, because devices are constrained by the bands that have been allocated for the given class.

\begin{figure*}
\begin{center}
\includegraphics[height=8cm]{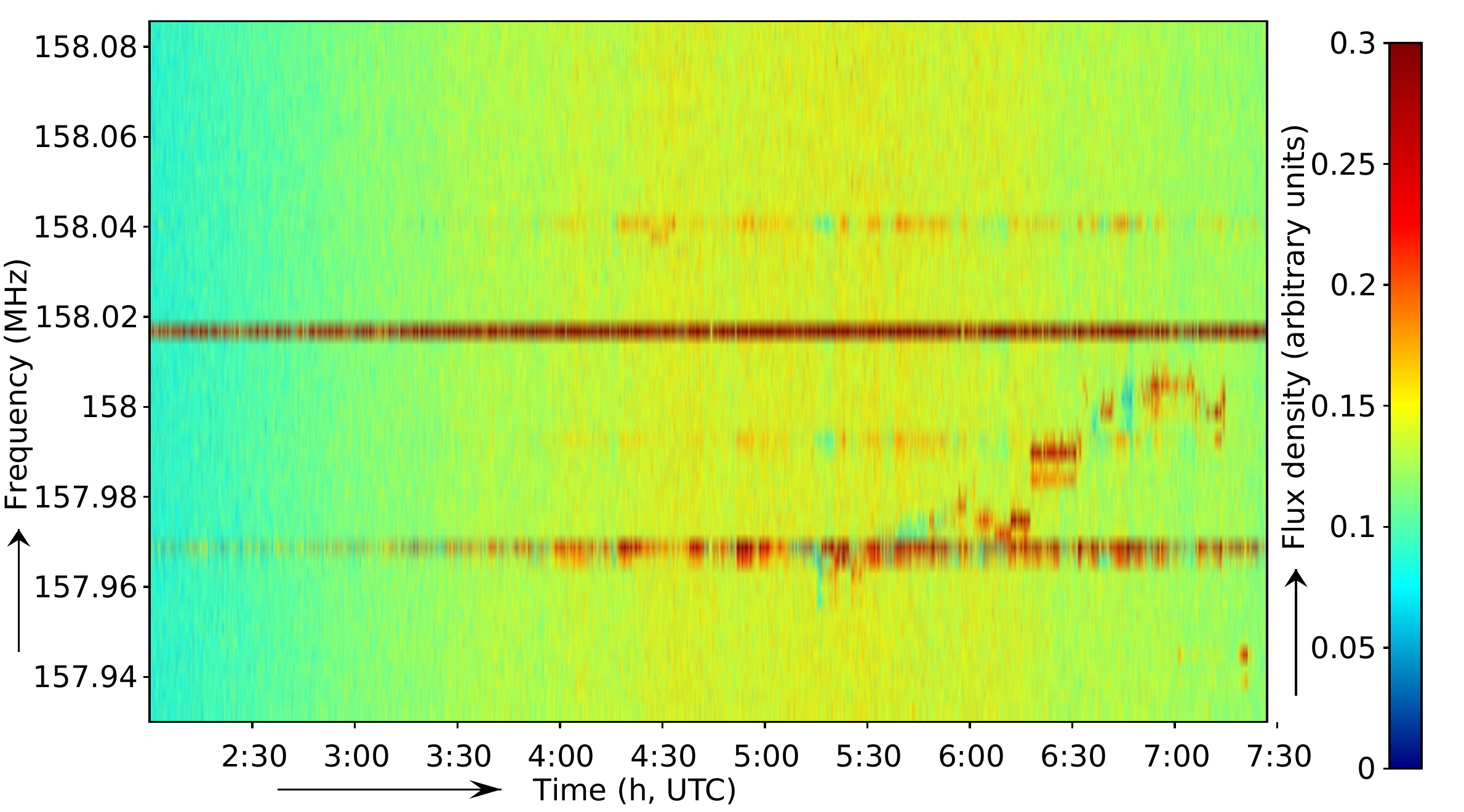}
\caption{A dynamic spectrum of a small part of an observation. The features with significantly higher values are caused by interference. Some of these have a constant frequency, while others are more erratic.}
\label{fig:dist-rfi-example}
\end{center}
\end{figure*}

In time-frequency space, interfering sources can have complex structures. They can also be intermittent and different sources might overlap in time-frequency space. An example of interfering sources can be seen in Fig.~\ref{fig:dist-rfi-example}, which shows raw visibility data of one baseline of a LOFAR observation in a dynamic spectrum. Because many sources change over time, are repetitive or affect multiple channels, many sources produce multiple unconnected features in time-frequency space. It is often not clear what constitutes a single interfering source, hence it is hard to count individual sources. Instead, we will count the number of times a given brightness occurs in time-frequency space. This --- as well as many other effects --- will of course influence the distribution. If sources overlap in the time-frequency space, the situation is somewhat similar to the case where multiple unresolved celestial radio sources in the reception pattern of a telescope only allow observation of a sum of sources. However, in that case it is still possible to validate radio source models by comparing $\log N$ -- $\log S$ histograms \citep{scheuer-analysing-faint-stars}.

It is common knowledge that in a uniform Euclidian Universe source counts behave like power-law distributions. The differential source-count distribution for sources on a flat surface is a power law with -2 exponent. We will derive this expected intrinsic source distribution for interfering radio sources. After that, we will analyse the issues that arise when measuring the distribution by counting samples.

In every dynamic spectrum we can measure the number of times that the flux density is within a particular range. Dividing this quantity by the total number of samples yields the relative number of events as a function of intensity. We will refer to this quantity with the term ``rate density''. We will now start by estimating the rate density function of ground-based interfering sources. Consider an interfering point source of strength $I$ that denotes the transmitting power normalized by the observational channel resolution (e.g., measured in W/Hz). This source is observed by an interferometer that consists of two antennas or stations with gains $g_1,g_2$, which include all instrumental effects. The antennas are located at distances $r_1,r_2$ from the source. The interferometer will record an apparent instantaneous strength $\mathcal{S}$ of
\begin{equation}
 \mathcal{S}(r_1, r_2) = I \frac{g_1 g_2}{4\pi r_1 r_2},
\end{equation}
with (real-valued amplitude) gains $g_1,g_2>0$ and $r_\textrm{L} > r_1,r_2 > 0$. Here, $r_\textrm{L}$ is a limiting distance, which will be well below the diameter of the Earth. The formula represents a spherically propagating wave in free space. We will limit our analysis to cross-correlated antennas; the auto-correlations will be ignored.

We assume that the source observed is fully coherent, but a possible de-coherence factor can be absorbed in the gains. Due to the small bandwidth of most interfering sources, most RFI will be received coherently, because of the narrow-band condition. With a frequency resolution $\Delta\nu=$ 0.76~kHz, the narrow-band condition $\Delta \nu \ll (2\pi \tau)^{-1}$ with correlation delay $\tau$ will hold for baselines up to a few km, because it holds as long as the baseline length is significantly less than $\Delta x = c (2 \pi \Delta \nu)^{-1} \approx$ 50~km. Because the velocity resolution of LOFAR is 1.5~km/s at 150~MHz, and larger at lower frequencies, a Doppler frequency shift due to movement of the source will only be significant if its velocity is at least 1.5~km/s relative to the antennas. Since the relative velocities towards different antennas in the array will be similar for such high-velocity transmitters (i.e., satellites), there will be hardly any decorrelation because of Doppler shifting.

Although two antennas do not necessarily observe the same RFI sources, for source-count analysis we can treat the interferometer geometrically as a single point, as both antennas will see the same distribution. Then, we can express the received amplitude $\mathcal{S}$ for a given distance $r$ and interferometric gain $g = g_1 g_2$ as
\begin{equation} \label{eq:amplitude-fall-off-in-free-space}
 S(r) = \frac{Ig}{4\pi r^2}.
\end{equation}

Next, we assume that all RFI sources have equal constant strength $I$ and follow a uniform spatial distribution in the local two-dimensional horizontal plane. These assumptions are obviously simplications, but we will address these later. Using these assumptions, we can express the expected inverse cumulative rate density of sources at distance $r$ as
\begin{equation} \label{eq:inverse-cumulative-distance-function}
F_{\textrm{distance} \ge r}(r) = N - \rho \pi r^2,
\end{equation}
with $N$ the total number of sources and for some constant $\rho$ that represents the number of sources per unit area. The cumulative number of sources $F_{\textrm{amplitude} \le S}$ that have an amplitude of at most $S$ can be calculated from this with
\begin{align} \label{eq:cumulative-rfi-distribution}
 F_{\textrm{amplitude} \le S}(S) = & F_{\textrm{distance} \ge r}(\mathcal{R}(S))
 = N-\frac{\rho Ig}{4S}
\end{align}
where $\mathcal{R}(S)=\mathcal{S}^{-1}$, the inverse of $\mathcal{S}$, i.e., the function that returns the distance $r$ for a given amplitude $S$. Finally, the rate density can be calculated by taking the derivative,
\begin{align} \label{eq:two-dimensional-distribution}
 f_S(S) = & \frac{dF_{\textrm{amplitude} \le S}}{dS} = \frac{\rho Ig}{4S^2}.
\end{align}

Therefore, if we plot the histogram of the RFI amplitudes in a $\log$-$\log$ plot, we expect to see a power law with a slope of $-2$ over the interval in which the RFI sources are spread like uniform sources on a two-dimensional plane.

\subsection{Propagation effects}
So far, we have assumed that the electromagnetic radiation propagates through free space, resulting in an $r^{-2}$ fall-off. In reality, the radiation will be affected by complicated propagational effects due to the surface of the Earth. A commonly used propagation model is the empirical model determined by \citet{okumura-propagation-model}, which was further developed by \citet{hata-propagation-loss}. Hata gives the following analytical estimate for $L_p$, the electromagnetic propagation loss between two ground-based antennas:
\begin{align} 
\notag                L_p & = & 69.55 + 26.16 \log_{10} f_c - 13.82 \log_{10} h_b - \\
\label{eq:hata-model} & &  a(h_m) + (44.9 - 6.55 \log_{10} h_b) \log_{10} r, 
\end{align}
where $L_p$ the loss in dB; $f_c$ the radiation frequency in MHz; $h_b$ the height of the transmitting antenna in meters; $h_m$ the height of the receiving antenna in meters; $r$ the distance between the antennas in meters; and $a(h_m)$ a correction factor in dB that corrects for the height of the receiving antenna and the urban density. \citeauthor{hata-propagation-loss} found this model to be representative for frequencies $f_c \sim$~150--1500~MHz, with transmitter heights $h_b \sim$~30--200~m, receiver heights $h_m \sim$~1--10~m and over distances $r\sim$~1--20~km.

Converting from a subtracted term in decibels to a flux density factor $L_S$ results in
\begin{equation} \label{eq:propagation-loss-factor}
 L_S = \frac{1}{10} 10^{L_p} = \zeta r^\eta,
\end{equation}
with $\eta$ and $\zeta$ given by
\begin{align} 
 \label{eq:zeta-definition} \eta & = 4.49 - 0.655 \log_{10} h_b,\\
 \label{eq:eta-definition}  \zeta & = \frac{f_c^{2.616}}{h_b^{1.382}} - 10^{6.955-\frac{1}{10} a(h_m)}.
\end{align}
Note that according to Hata's model, the exponent of the power law $\eta$ depends only on the height of the transmitting antenna, i.e., it is independent of frequency, receiver height and urban density. To find the rate density function $f_p$ that considers propagation effects, one can replace $S(r)$ in Eqs.~\eqref{eq:cumulative-rfi-distribution} and \eqref{eq:two-dimensional-distribution} with one that includes the propagation effects,
\begin{equation}
 S(r) = \frac{Ig}{4\pi \zeta r^\eta}.
\end{equation}
The resulting rate density function $f_p$ is
\begin{align} \label{eq:frequency-density-with-propagation}
f_p(S)
= & \frac{d}{dS} \left[ N - \rho \pi \left( \frac{Ig }{4\pi\zeta S} \right)^{2/\eta} \right]
= \frac{\rho 2\pi}{\eta S} \left( \frac{Ig}{4\pi\zeta S} \right)^{2/\eta}.
\end{align}
Consequently, due to non-free-space propagation effects, the observed log-log histogram is predicted to have a $-(\frac{2}{\eta}+1)$ slope. By substituting $\eta$, one finds
\begin{equation}
\textrm{slope}(h_b) = \frac{1}{0.3275 \log_{10} h_b - 2.245} - 1.
\end{equation}
\begin{figure}
\begin{center}\hspace{-1cm}\includegraphics[width=6cm]{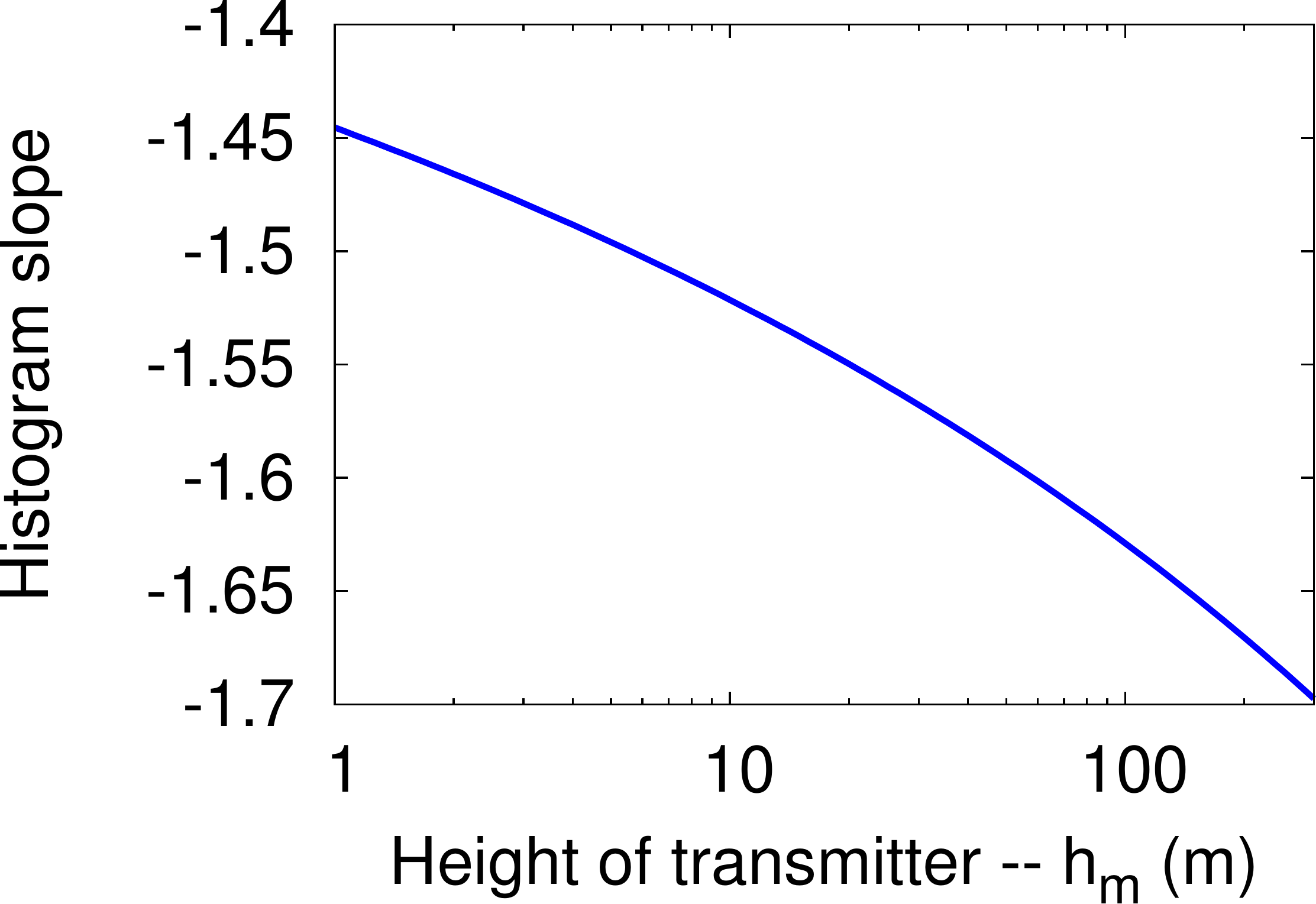}
\caption{Effect of transmitter height on the slope of a log-log histogram. According to Hata's model, this is valid for the range 30--200 m. The trend of the slope will not continue indefinitely when increasing the height further. Instead it will converge to a $-2$ slope, which corresponds to free-space propagation.}
\label{fig:plot-transmitter-height}
\end{center}
\end{figure}
This yields estimated distribution slopes of $-1.57$ and $-1.67$ for $30$ m and $200$ m high transmitters respectively. In Figure~\ref{fig:plot-transmitter-height}, the slope value is plotted as a function of the transmitter height, including extrapolated values for transmitter heights down to 1 m.

We note that a uniform distribution of meteors or aircrafts which reflect free-space propagating RFI can create a power-law distribution with a similar slope: a uniform two-dimensional distribution of reflecting sources will create a $-1.5$ slope, while a uniform three-dimensional distribution will create a $-1.75$ slope. With brightness-distribution analyses one can therefore not distinguish between transmitters affected by Hata's propagation model and reflectors affected by free-space propagation. Reflected RFI might become relevant at lower amplitude levels.

\subsection{Thermal noise contribution} \label{sec:histogram-noise}
The full measured distribution will consist of the power-law distribution combined with that of the thermal noise and the celestial signal. For now, we will ignore the contribution of the celestial signal, as its contribution to the amplitude distribution will be minimal when observing fields without strong celestial sources. For example, the strongest apparent celestial source in the NCP EoR field is around 5 Jy \citep{ncp-eor-yatawatta}.
The standard deviation of the noise, however, is around 100~Jy on highest LOFAR resolutions, and will have a larger contribution on the histogram.

\begin{figure}
\begin{center}\hspace{-5mm}\includegraphics[width=8cm]{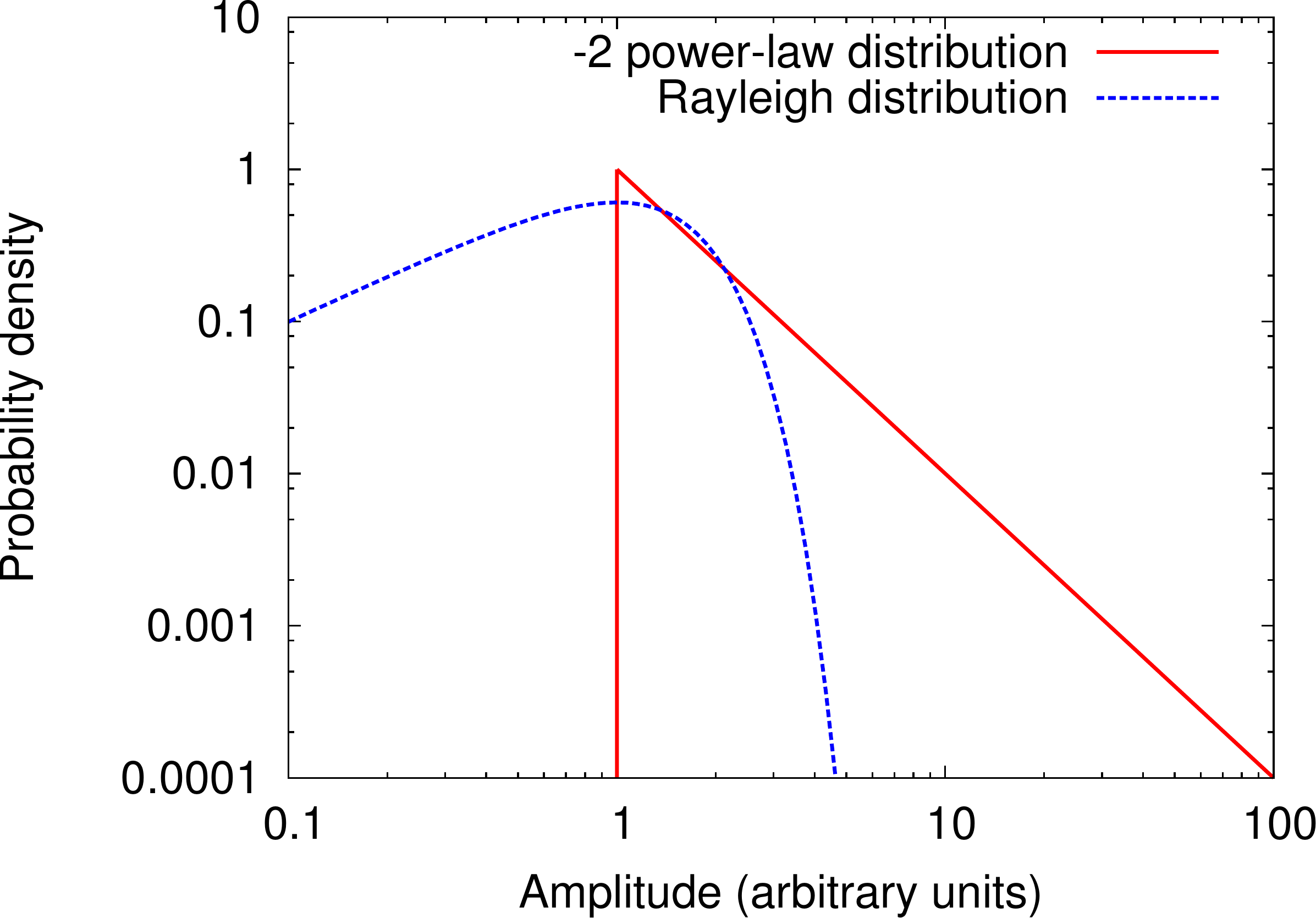}
\caption{The Rayleigh and power-law distributions in a log-log plot. The power-law distribution (Eq.~\eqref{eq:two-dimensional-distribution}) has a constant slope of -2 over the range it is defined. The slope of the Rayleigh distribution in the limit of the origin is 1. Its maximum occurs where the amplitude value equals its mode $\sigma$, which is 1 in this example. For higher amplitudes, its slope decreases exponentially.}
\label{fig:rayleigh-and-rfi-distributions}
\end{center}
\end{figure}

The real and imaginary components of the noise in the cross-correlations are independent and identically Gaussian distributed with zero mean and equal variance. Consequently, an amplitude $x$ will be Rayleigh distributed \citep[\S6-2]{papoulis-stochastic-processes}: 
\begin{equation}\label{eq:rayleigh-formula}
f_\textrm{noise}(x) =
\begin{cases}
\frac{x}{\sigma^2} e^{\frac{-x^2}{2\sigma^2}} & x > 0, \\
0 & \textrm{otherwise.}
\end{cases}
\end{equation}
Because most of the samples will be unaffected by RFI, this will be the dominating distribution. The Rayleigh distribution is plotted together with the -2 power-law distribution of Eq.~\eqref{eq:two-dimensional-distribution} in Fig.~\ref{fig:rayleigh-and-rfi-distributions}.

\begin{figure}
\begin{center}\hspace{-2mm}\includegraphics[width=8.5cm]{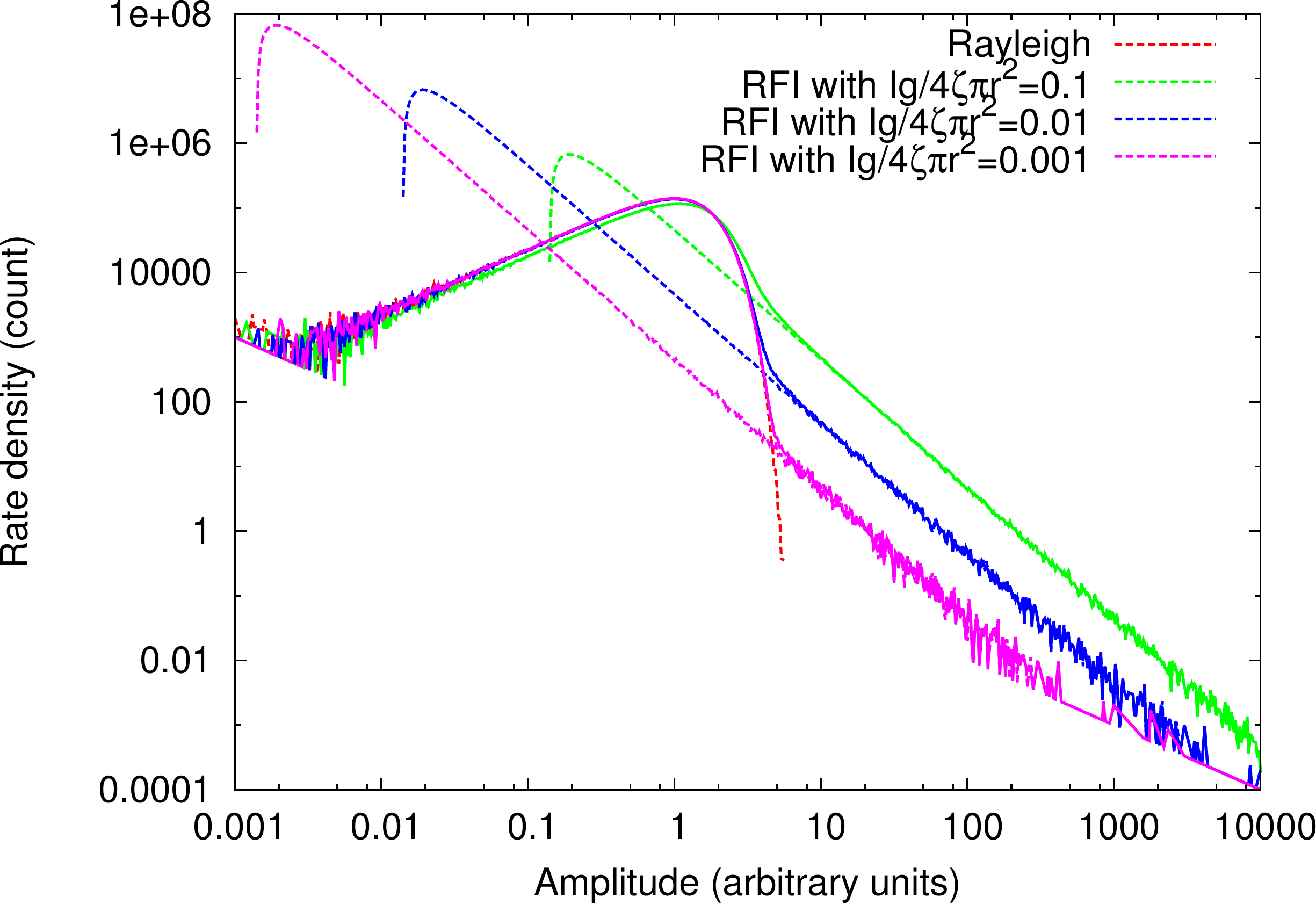}
\caption{Histograms of simulated samples that all have a contribution of noise and RFI. Various settings of the parameters were used, and samples were drawn as described in Eq.~\eqref{eq:drawing-rfi-with-propagation}. Solid lines: the combined distributions, dashed lines: the power-law distributions before mixing. }
\label{fig:rayleigh-and-rfi-combined}
\end{center}
\end{figure}

So far, these are the expected histograms for pure noise and pure RFI that propagates through free space. However, the measured distribution is a mixture of the two. Analytic derivation of the corresponding mixed amplitude distribution is not trivial, but the distributions can easily be estimated by drawing complex samples from the two distributions and calculating and counting the amplitudes. A sample can be drawn from the RFI distribution by integration, scaling and inversion of the rate density function in Eq.~\eqref{eq:frequency-density-with-propagation}. To invert the cumulative function, one needs to assume that there are no sources beyond some limiting distance $r_L$. With this assumption, a single complex RFI contaminated sample $S_{\textrm{RFI}}$ can be sampled with:
\begin{equation} \label{eq:drawing-rfi-with-propagation}
S_{\textrm{RFI}} \leftarrow \frac{I g}{4\pi \zeta x_u^{\eta/2} r_L^\eta} e^{i 2 \pi y_u}.
\end{equation}
Here, $S_{\textrm{RFI}}$ is a new complex RFI sample that follows a power-law distribution; $\eta$ and $\zeta$ are defined in Eqs.~\eqref{eq:zeta-definition} and \eqref{eq:eta-definition}; $I$ is the average intrinsic strength; $g$ is the gain of the instrument; $0 < x_u,y_u \le 1$ are two independently drawn uniformly distributed samples; and $r_L$ is the maximum distance of visible sources. A sample $S$ that is contaminated by both RFI and noise can be drawn with
$S \leftarrow v_n + w_n i + S_{\textrm{RFI}}$, with $v_n, w_n \sim N(\mu=0; \sigma)$. An example of distribution curves of $S$ for $\eta=2$ and various settings of $Ig/4\pi \zeta r_L^2$ is given in Fig.~\ref{fig:rayleigh-and-rfi-combined}.

\subsection{Parameter variability} \label{sec:parameter-variability}
In reality, the parameters $\rho$, $I$ and $g$, which are the RFI source density per unit area, RFI source strength and instrumental gain respectively, will not be constant, but can change over time and frequency. Therefore, they are stochastic variables. However, since each specific value for these parameters produces a power law, the combined distribution will still show a power law, as long as the parameters follow a distribution that is steep at high amplitudes (in $\log$--$\log$ space), such as a Gaussian or uniform distribution.

One instrumental effect that is absorbed in $g$ is the frequency response of the instrument, i.e., the antenna response in combination with the band-pass of the analogue and digital filters. Because the data that are analysed in Sect.~\ref{sec:dist-results} have initially not been band-pass calibrated, the instrumental response is not uniform over frequency. We determined that the gain variation due to the band-pass is about one order of magnitude for the low-band antennas (LBA, 30.1--77.5 MHz) and about a factor of two for the high-band antennas (HBA, 115.0--163.3 MHz). The frequency dependency of the gains due to the band-pass will consequently smooth the data in the brightness histogram in horizontal direction by one order of magnitude or less.

Another effect that is absorbed in $g$, is the beam of the instrument. At the point of writing, LOFAR beam models are still being developed and are not yet well parametrized near the horizon. It is likely that most RFI sources are observed at the edges of the beam. Nevertheless, most sources will be observed with similar gains (within one order of magnitude), and it can be expected that the beam will have a limited effect on the histogram properties of an observation. It is therefore comparable with the effect of the frequency response.

The stochastic nature of $I$, that is caused by the spread of transmitters with different intrinsic strengths, might also have an effect on the $\log N$ -- $\log S$ histograms. It is unlikely that $I$ follows a uniform or Gaussian distribution, because the distribution will contain few strong transmitters (such as radio stations) and many weak transmitters (such as remote controls). Therefore, variable $I$ might follow a power-law distribution by itself. It is likely that strong transmitters transmit more on average, and therefore contaminate more samples as well. High-power transmitters, such as radio stations, have a typical equivalent isotropically radiated power (EIRP) in the order of 10--100~kW. Low-power transmitters, such as remote controls, transmit with an order of 100~mW or even less. Therefore, these devices have a spread of around 6 orders of magnitudes in power. As long as the exponent of the power law of $I$ is less steep (i.e., less negative) compared to the power law caused by the spatial distribution, the spatial distribution will dominate the histogram at high amplitudes. With a spatial $-1.5$ power law and the given transmitting powers, the low-power transmitters should contaminate a factor of $10^9$ more samples compared to the high-power transmitters to dominate the high-amplitude distribution, which is unlikely. Therefore, it is likely that the spatial distribution will dominate the power law in the histogram. Otherwise, a turn-over point should be visible in the histogram.

From Eq.~\eqref{eq:frequency-density-with-propagation} it can be seen that the $\rho$, $I$ and $g$ parameters have the same effect of scaling the power-law distribution, and do not change its shape or slope. Therefore, with distribution analyses one can e.g. not determine whether the distribution is dominated by low-power sources within the horizon or by scattered signals from over the horizon. The horizon of an antenna is estimated with $\sqrt{2hr}$ \citep{bullington-propagation}, with $r$ the radius of the Earth and $h$ the height of the antenna. For LOFAR, the horizon is at about 5~km.

\section{Methods} \label{sec:distribution-methods}
In this section we will briefly discuss how the histograms are created, how the slope of the underlying RFI distribution is estimated and show how to constrain some of the intrinsic RFI parameters.

\subsection{Creating a histogram} \label{sec:histogram}
While creating a histogram is trivial, it is important to note that it is necessary to have a variable bin size. This is mandated by the large dynamic range of the histogram that we are interested in. Therefore, we chose to have a bin size that increases linearly with the amplitude $S$, and the rate counts are divided by the bin size after counting. 

\subsection{Estimating $\sigma$ and slope parameters}
The mode $\sigma$ of the Rayleigh distribution is estimated by finding the amplitude with the maximum occurrences, i.e., the amplitude corresponding to the peak of the histogram. The slope is estimated using linear regression over a visually selected interval. We have validated that the slope does not significantly change by using a slightly different interval.

Fitting straight lines to the distribution curve in a log-log plot is not the most accurate way of estimating the exponent of a power-law distribution \citep{power-law-distribution}. However, because of our enormous sample size, which allows fitting the line over a large interval, the estimator will be sufficiently accurate for our purpose. Nevertheless, we will additionally calculate a maximum-likelihood estimator for comparison. The maximum-likelihood estimator for the exponent in a power-law distribution is given by the Hill estimator $\hat \alpha_H$ \citep{hill-estimator, power-law-distribution}, defined as:
\begin{equation} \label{eq:hill}
 \hat \alpha_H = 1 + N \left(\sum\limits_{i=1}^{N} \ln \frac{x_i}{x_\textrm{min}} \right)^{-1},
\end{equation}
with $N$ the number of samples and $x_i$ for $0 < i \le N$ the samples that follow a power law with lower bound $x_\textrm{min}$.

\begin{figure}
\begin{center}
\includegraphics[width=7.5cm]{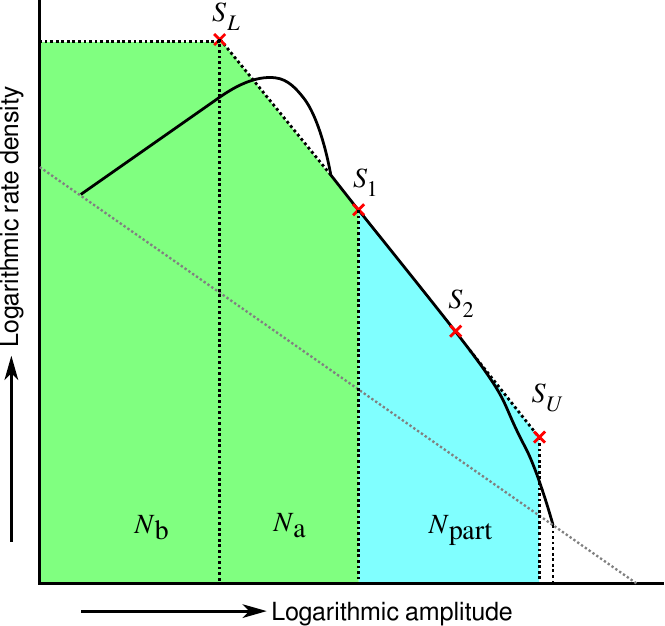}
\caption{Cartoon of how a constraint on the lower fall-off point of the power-law distribution can be determined. Note that the labelled areas are areas as occupied in a linear plot, i.e., the integration of the density function. Areas in a log-log plot are not linearly related to the integral. There are two ways to estimate the lower constraint $S_L$: (i) the areas $N_a$ and $N_\textrm{total}-N_\textrm{part}$ are equal if $Ig/r_L^\eta$ is constant during the observation, and (ii) if one assumes $Ig/r_L^\eta\sim$ uniform, then $N_a+N_b = N_\textrm{total}-N_\textrm{part}$.}
\label{fig:explanation-constraints}
\end{center}
\end{figure}

\subsection{Determining RFI distribution limits} \label{sec:rfi-distribution-constraints}
In this section we will show how to put upper and lower constraints on the power-law distribution. Assume that we have found a power law with exponent $\alpha$ and factor $\beta$ over an amplitude region $[S_1 ; S_2]$, resulting in the rate density function $h(S) = \beta S^\alpha$. $S_1$ and $S_2$ are selected by visual inspection of the histogram. Assume that the histogram contains $N_\textrm{part}$ (RFI) samples with amplitude $> S_1$, as sketched in Fig.~\ref{fig:explanation-constraints}, and that the effect of the Rayleigh component on the histogram $> S_1$ is negligible. The hypothetical upper limit $S_U$ of the distribution can be found by solving
\begin{equation} \label{eq:upper-limit-constrained}
 \int\limits_{S_1}^{S_U} h(S) dS = N_\textrm{part}.
\end{equation}
The observed histogram will break down beyond some amplitude because of several reasons: the samples are digitized with an analogue-to-digital converter (ADC) with limited range; we observe for a limited time and the rate count is discrete; and, under the assumption of a uniform spatial distribution of RFI transmitters, samples with very high amplitude would have to be produced by transmitters that are very close to the telescope. However, it is likely that the uniform spatial distribution of transmitters will break down at closer distances.

Solving Eq.~\eqref{eq:upper-limit-constrained} results in
\begin{equation} \label{eq:upper-limit}
S_U = \sqrt[\alpha+1]{\frac{\alpha+1}{\beta} N_\textrm{part} + S_1^{\alpha+1}}.
\end{equation}

One can estimate the lower limit $S_L$ in a similar way. This can be solved by assuming the area labelled $N_a$ in Fig.~\ref{fig:explanation-constraints} equals the number of samples to the left of $S_1$. The area labelled $N_b$ is assumed to be zero for now, which assumes the power law has a sharp cut-off on the left side, e.g., because of the curvature of the Earth. Solving $N_a = N_\textrm{total}-N_\textrm{part}$ results in
\begin{equation} \label{eq:lower-limit-1}
S_L = \sqrt[\alpha+1]{\frac{\alpha+1}{\beta}\left(N_\textrm{part} - N_\textrm{total} \right) + S_1^{\alpha+1}}.
\end{equation}

With the assumption that $Ig/r_L^\eta\sim$ a uniform distribution, the area labelled in Fig.~\ref{fig:explanation-constraints} as $N_b$ is also part of the RFI distribution, and a stronger constraint $\tilde S_L$ can be found, yielding
\begin{equation} \label{eq:lower-limit-2}
\tilde S_L = \sqrt[\alpha+1]{ - \frac{1}{\alpha} \left( \frac{\alpha+1}{\beta} \left(N_\textrm{part} - N_\textrm{total}\right) + S_1^{\alpha+1} \right) }.
\end{equation}
With estimates of $\alpha$, $\beta$, $S_L$ and $S_U$, one has obtained a parametrization of the RFI distribution. As was shown in \S\ref{sec:histogram-noise}, the left-most point where the power-law distribution falls off is $S_L = Ig / 4\pi \zeta r_L^\eta$. This value represents the apparent brightness of the RFI sources that are furthest away from the telescope. With a fully parametrized distribution of the effect of RFI sources, an empirical model for RFI effects can be made. Moreover, one can calculate $E(S_R)$, the expected apparent strength of RFI:
\begin{align} \label{eq:expected-value-rfi-def}
E(S_R) & = \frac{1}{N_{LU}}\int\limits_{S_L}^{S_U} \beta S^\alpha S dS
 = \frac{\beta}{N_{LU}} \left[ \frac{1}{\alpha+2} S^{\alpha+2} \right]_{S_L}^{S_U}
\end{align}
Here, $N_{LU}$ is the number of samples between $S_L$ and $S_U$ after normalizing for the bin size. Evaluating this results in
\begin{align} \label{eq:expected-value-rfi}
 E(S_R) & = \frac
{\left( S_U^{\alpha+2} - S_L^{\alpha+2}\right)\left(\alpha+1\right)} 
{\left(S_U^{\alpha+1} - S_L^{\alpha+1}\right) \left(\alpha+2 \right)}.
\end{align}
This is essentially the average flux density that is caused by RFI without using RFI detection or excision algorithms. $E(S_R)$ has the same units as $S_L$ and $S_U$, thus after calibration (see \S\ref{sec:calibration}) could be given in Jy. In practice, the increase of data noise after correlation is much less severe because of RFI flagging, which excises a part of the RFI. One can assume that all RFI above a certain power level is found by the detector. Since modern RFI detection algorithms can find all RFI that is detectable ``by eye'' \citep{post-correlation-rfi-classification}, this power level will be near the level of the noise mode. We use the AOFlagger for RFI detection, which will be described in Sect.~\ref{sec:dist-data}.

Another interesting parameter is $S_d$, the average lower limit of detected RFI. It can be calculated by finding the point on the distribution where the area under the distribution to the right of $S_d$ equals the real number (true positives) of RFI samples. Therefore, the limit is calculated similar to Eq.~\eqref{eq:lower-limit-1}, where the term $N_\textrm{part} - N_\textrm{total}$ needs to be replaced with $N_\textrm{RFI}$, which equals the total number of samples detected as RFI minus the false positives. In \citet{lofar-radio-environment} the false-positives rate for the AOFlagger is estimated to be 0.5\%.

Finally, $E(S_\textrm{leak})$, which is the expected value of leaked RFI not detected by the flagger, can be calculated by replacing $S_U$ with $S_d$ in the numerator of Eq.~\eqref{eq:expected-value-rfi} and subtracting the removed number of samples from the total of number of samples. Assume that a fraction of $\kappa$ samples are not detected as RFI and $1-\kappa$ have been detected as RFI, then
\begin{align} \label{eq:expected-value-leaked-rfi}
E(S_\textrm{leak}) & = \frac{1}{\kappa N_{LU}}\int\limits_{S_L}^{S_d} \beta S^\alpha S dS
= \frac{\left( S_d^{\alpha+2} - S_L^{\alpha+2} \right) \left( \alpha+1 \right) } {\kappa \left( S_U^{\alpha+1} - S_L^{\alpha+1}\right) \left( \alpha+2 \right)}.
\end{align}
This is the average contribution that leaked RFI will have on a single sample. It has the same units as the parameters $S_L$, $S_U$ and $S_d$. Typical values for $\kappa$ are $0.95$--$0.99$.

\subsection{Calibration} \label{sec:calibration}
We can assign flux densities to the horizontal axis of the histogram by using the system equivalent flux density (SEFD) of a single station. The current LOFAR SEFD is found to be approximately 3400~Jy for the HBA core stations and 1700~Jy for the remote stations in the frequency range from $125$--$175$ MHz. For all Dutch LBA stations, in the frequency range $40$--$70$ MHz the SEFD is approximately 34,000~Jy. The standard deviation $\sigma$ in the real and imaginary values is related to the SEFD with 
\begin{equation}
 \sigma = \frac{\textrm{SEFD}}{\sqrt{2 \Delta \nu \Delta t}},
\end{equation}
where $\Delta \nu$ is the bandwidth and $\Delta t$ is the correlator integration time. The standard deviation will appear as the mode of the Rayleigh distribution. By fitting a Rayleigh function with fitting parameter $\sigma$ to the distribution, one finds the corresponding flux density scale.

RFI sources will enter through the distant sidelobes of the station beams from many unknown directions. Moreover, models for the full beam are often hard to construct. Therefore, we will not try to calibrate the beam, and the flux densities in the histogram are apparent quantities. Consequently, we will not be able to say something about the true intrinsic power levels of RFI sources.

\subsection{Error analysis}
An estimate for the standard deviation of the slope estimator $\hat \alpha$ can be found by calculating $\textrm{SE}(\hat \alpha)$, the \emph{standard error} of $\hat \alpha$. The standard error of the slope of a straight line \citep[pp. 32--35]{acton-analaysis-of-straight-lines} is given by
\begin{equation} \label{eq:stderr-slope}
 \textrm{SE}(\hat \alpha) = \sqrt{\frac{SS_{yy}-\hat\alpha SS_{xy}}{\left(n - 2\right) SS_{xx}}},
\end{equation}
where $SS_{xx}$, $SS_{xy}$ and $SS_{yy}$ are the sums of squares, e.g., $SS_{xy}=\sum_{i=1}^n (x_i - \bar x) (y_i - \bar y)$ and $n$ is the number of samples. However, we found that this is not a representative error in our case, because the errors in the slope are not normally distributed. Therefore, we also introduce an error estimate $\epsilon_\alpha$ that quantifies a normalized standard deviation of the slope over the range. This error is formed by calculating the slope over $n_\alpha$ smaller sub-ranges in the histogram, creating $n_\alpha$ estimates $\alpha_i$. If the errors in $\alpha_i$ are normally distributed with zero mean, an estimate of the variance of $\hat \alpha$ can be calculated with
\begin{equation}
 \epsilon_{\hat\alpha} = \sqrt{\frac{\sum \left( \alpha_i - \hat\alpha \right)^2 }{n^2_\alpha - n_\alpha}}.
\end{equation}
This estimate is slightly depending on the number of sub-ranges that is used, $n_\alpha$, because the errors are not Gaussian distributed, but we found that $\epsilon_{\hat\alpha}$ is more representative than the standard error of $\hat\alpha$. 

The standard error of the Hill estimator of Eq.~\eqref{eq:hill} is \citep{power-law-distribution}
\begin{equation} \label{eq:stderr-hill}
 \textrm{SE}(\hat \alpha_H) = \frac{-\alpha - 1}{\sqrt{n}} + \mathcal{O}(\frac{1}{n}).
\end{equation}
Because the number of samples is very large ($>10^{11}$), the $\mathcal{O}$-term will be very small. Therefore, we will calculate the quantity without this term.

\section{Data description} \label{sec:dist-data}
We have analysed the distributions of two data sets. Both data sets are 24-h LOFAR RFI surveys and are extensively analysed in \citet{lofar-radio-environment}. We refer to \citet{lofar-overview-2013} for a full description of the capabilities of LOFAR. The analyses will cover only Dutch stations. Each Dutch station consists of 96 dipole low-band antennas (LBA) and one or two fields totalling 48~tiles of 4x4 bow-tie high-band antennas (HBA). The core area of LOFAR is located near the village of Exloo in the Netherlands, where the station density is at its highest. The six most densely packed stations are on the Superterp, an elevated area surrounded by water situated 3~km North of Exloo. A radio-quiet zone of 2~km around the Superterp has been established, but is relatively small and households exist within 1~km of the Superterp. With the help of the spectrum allocation registry, the most-obvious transmitters can easily be identified and ignored in LOFAR data \citep{lofar-radio-environment}. However, many interfering sources have an unknown origin.

In the two data sets, we have used the correlation coefficients of cross-correlated stations, i.e., the raw visibilities. In one data set, the low-band antennas (LBA) were used and the frequency range 30.1--77.5~MHz was recorded, while in the other the high-band antennas (HBA) were used to record the frequency range 115.0--163.3~MHz. More stations were used in the LBA set. The specifications of the two sets are listed in Table~\ref{table:dist-data-specs}. The stations that have been used are geometrically spread over an area of about 80~km and 30~km in diameter at maximum for the LBA and HBA sets respectively. For EoR detection experiments, the HBA are more important than the LBA, because they cover the frequency range of the redshifted EoR signal.

Although we have used Hata's model to estimate the RFI log-log histogram slope, our frequency range falls partly outside the frequency range over which Hata's model has been verified. However, according to Hata's model the observing frequency does not influence the power-law exponent in the frequency range 150--1500 MHz, thus it can be assumed the exponent will at least not significantly differ over the HBA range.

To detect RFI, the AOFlagger \citep{LOFAR-RFI-pipeline} is used. This RFI detector estimates the sky contribution by iteratively applying a high-pass filter to the visibility amplitudes of a single baseline in the time-frequency plane. Subsequently, it flags line-shaped features with the SumThreshold method, which is a combinatorial threshold method \citep{post-correlation-rfi-classification}. Finally, the scale-invariant rank operator, a morphological technique to search for contaminated samples, is applied on the two-dimensional flag mask \citep{scale-invariant-rank-operator}.

Because the AOFlagger detector is partly amplitude-based, it is likely that low-level RFI will leak through the detector. Since it is also low-level RFI we are interested in, we will analyse unflagged data and the RFI classified data.

\begin{table}
\caption{Data set specifications}\label{table:dist-data-specs}
\begin{center}
\begin{tabular}{lrr}
                    & \textbf{LBA set}& \textbf{HBA set} \\
\hline
\hline
Observation date    & 2011-10-09      & 2010-12-27 \\
Start time          & 06:50 UTC       & 0:00 UTC \\
Length              & 24 h           & 24 h \\
Time resolution     & 1 s             & 1 s \\
\hline
Frequency range     &  30.1--77.5 MHz & 115.0--163.3 MHz\\
Frequency resolution & 0.76 kHz    & 0.76 kHz \\
Number of stations  &  33           & 13 \\
Total size          & 96.3 TB        & 18.6 TB \\
\hline
Field               & NCP             & NCP \\
Amount of RFI detected & & \\
by the AOFlagger    & 1.77\%       & 3.18\% \\
\hline
\hline
\end{tabular}
\end{center}
\end{table}

\begin{figure*}
\begin{center}
\includegraphics[width=12cm]{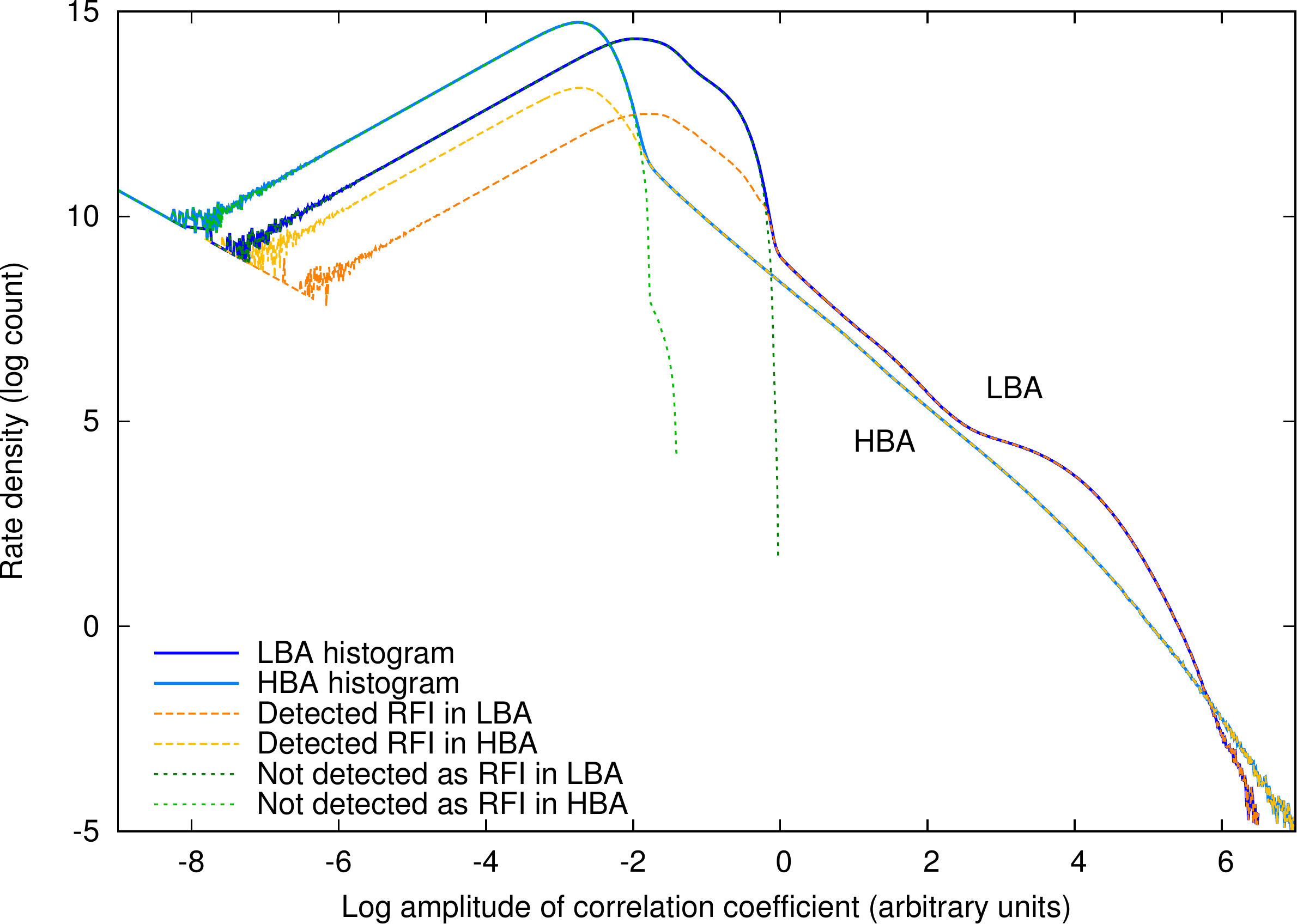}
\caption{The histograms of the two data sets before band-pass correction and flux calibration.}
\label{fig:Histograms-raw}
\end{center}
\end{figure*}

\section{Results} \label{sec:dist-results}
In this section we present the histograms of the LBA and HBA sets and the results that were obtained by applying the methodology discussed in Sect.~\ref{sec:distribution-methods}. 

\begin{figure}
\begin{center}
\includegraphics[width=8cm]{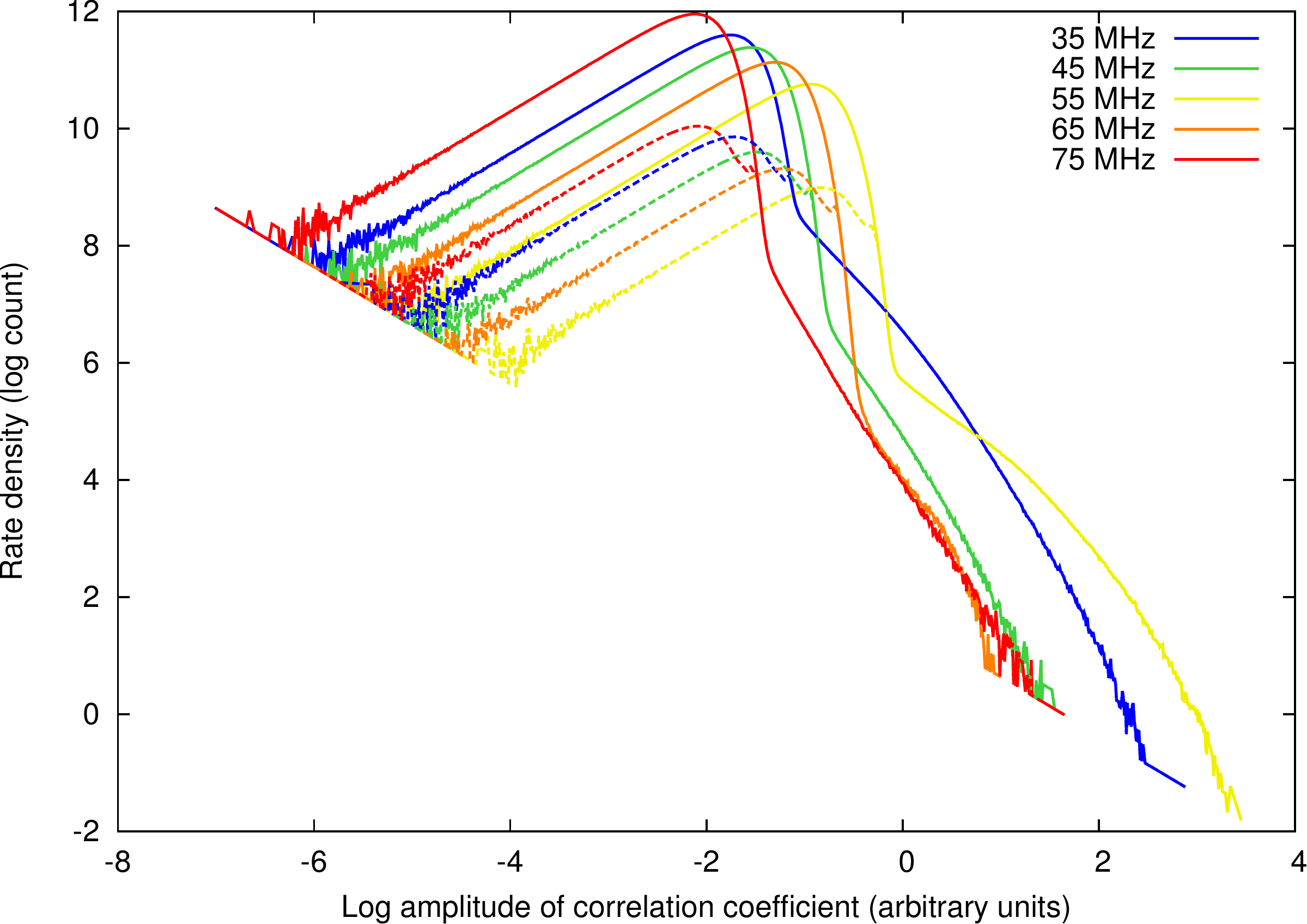}
\caption{Histograms for $5$ different $0.2$ MHz LBA sub-bands without band-pass correction and flux calibration. The continuous lines represent the data before RFI flagging. The dashed lines are the histograms of the samples that have been classified as RFI.}
\label{fig:plot-dist-per-frequency-LBA}
\end{center}
\end{figure}

\subsection{Histogram analysis}
Fig.~\ref{fig:Histograms-raw} shows the histograms with logarithmic axes for the LBA and HBA sets. In both sets, it is clear that at least one component with a Rayleigh and one with a power-law distribution have been observed. The left part of the histogram matches the Rayleigh distribution well up to the mode of the distribution. The bulge around the mode of the LBA histogram is wider due to the larger effect of the antenna response, i.e., variability of $g$ as discussed in \S\ref{sec:parameter-variability}. As can be seen in Fig.~\ref{fig:plot-dist-per-frequency-LBA}, the Rayleigh-bulges of individual sub-bands are not that wide, but they are not aligned because of the differing noise levels. 

\begin{figure}
\begin{center}
\includegraphics[width=8cm]{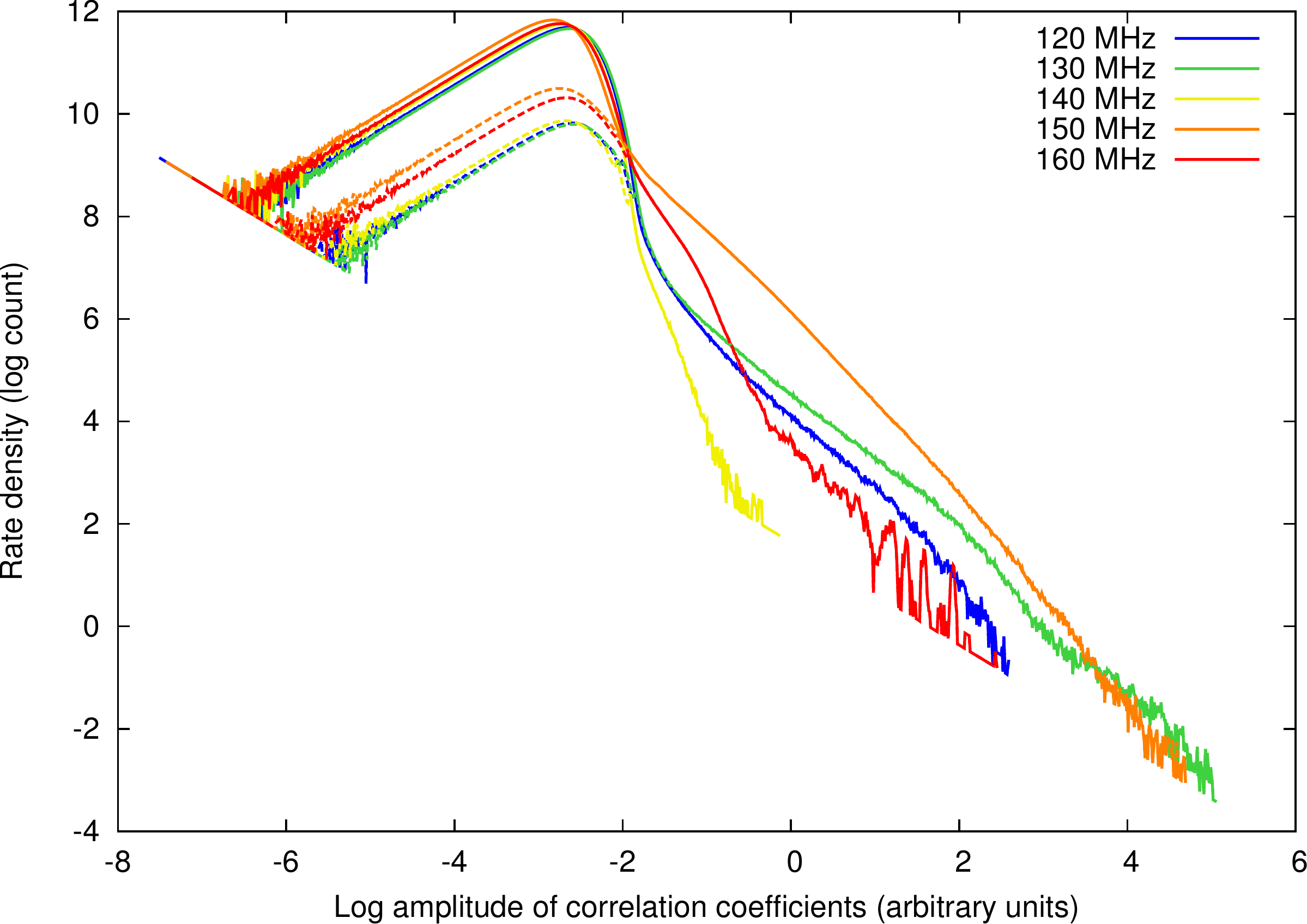}
\caption{Histograms for $5$ different $0.2$ MHz HBA sub-bands without band-pass correction and flux calibration. The continuous lines represent the data before RFI flagging. The dashed lines are the histograms of the samples that have been classified as RFI.}
\label{fig:plot-dist-per-frequency-HBA}
\end{center}
\end{figure}

It is to be expected that the RFI-dominated part of the distributions at different frequencies will reflect the underlying RFI source populations. Both Figs.~\ref{fig:plot-dist-per-frequency-LBA} and \ref{fig:plot-dist-per-frequency-HBA} show that the power-law part of the distributions are very different for different sub-bands. Nevertheless, combining the data of all the sub-bands results in reasonably stable power-law distributions. The variation could be caused by the different power-law exponents that source populations at different frequencies might have. It could also be caused by a differing number of transmitters. In that case, the underlying power law might not always be apparent, because not enough samples are combined. By making distributions over different frequency ranges, we have verified that the power law is not dominated by a few obvious and known sources.

\begin{figure*}
\begin{center}
\includegraphics[width=13cm]{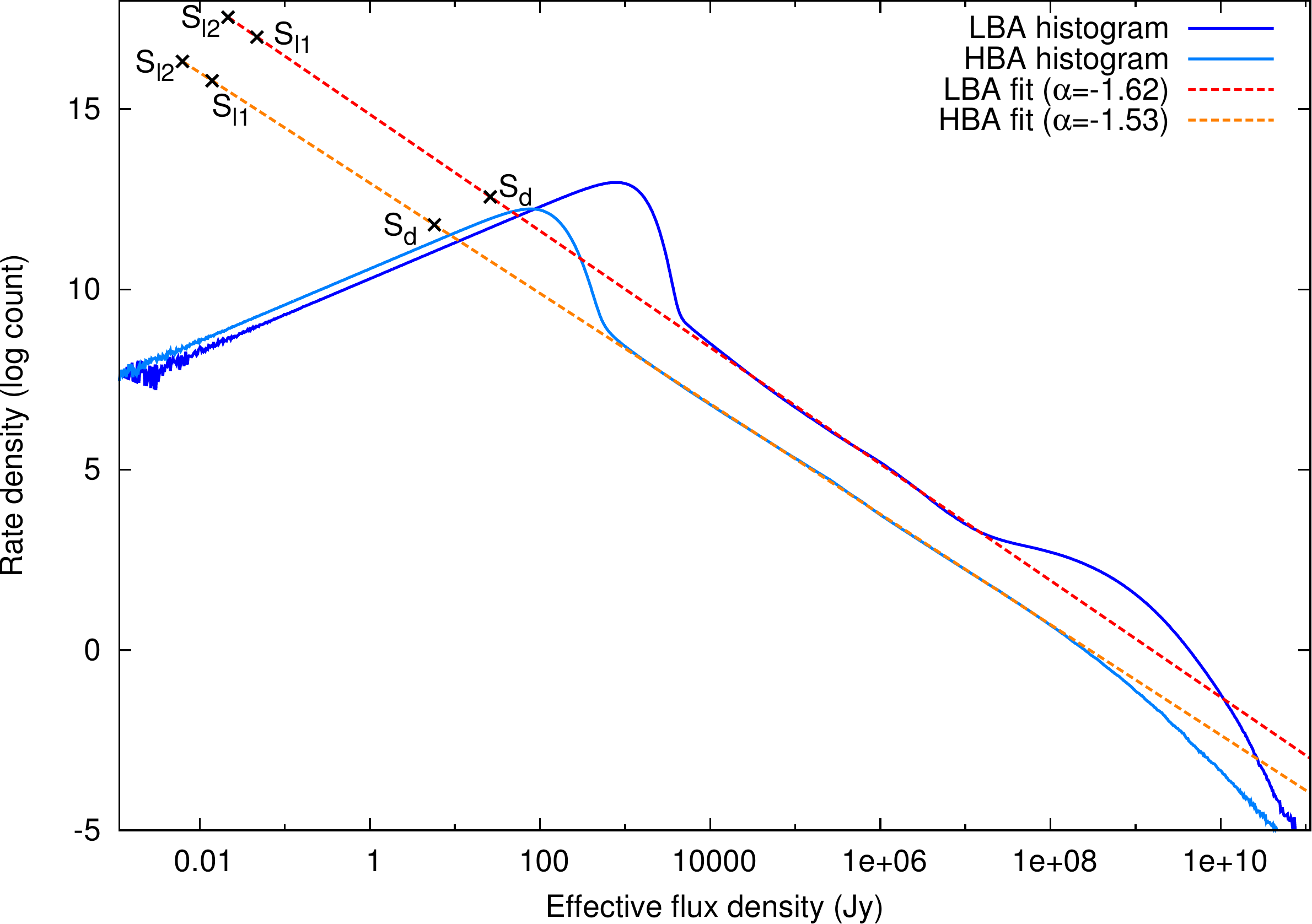}
\caption{Observed LBA distribution after band-pass correction and flux calibration. $S_{l1}$ and $S_{l2}$ denote the limits of the distribution with a sharp lower cut-off (Eq.~\eqref{eq:lower-limit-1}) and uniform lower limit (Eq.~\eqref{eq:lower-limit-2}), $S_d$ is the average lower limit of RFI that is detected.}
\label{fig:histogram-passband-corrected}
\end{center}
\end{figure*}

To make sure that the antenna response does not influence the result of the slope, we have also analysed the curves after a simple band-pass calibration. This was performed by dividing each sub-band by its standard deviation after RFI excision. Because the standard deviation of the distribution might be affected by the RFI tail of the distribution, we compare the two histograms to make sure the power-law distribution is not significantly changed. The resulting histograms are shown in Fig.~\ref{fig:histogram-passband-corrected}. This procedure makes the bulge of the LBA histogram similar to the bulge of a Rayleigh curve and extends the power-law part. Nevertheless, it does not change the log-log slope of the power law in either histograms. This validates that the variable gain that is caused by the antenna response does not change the observed power law. Consequently, it can be expected that other stochastic effects, such as the intrinsic RFI source strength and the beam gain due to a differing direction of arrival, will similarly not affect the power law. Because the band-pass corrected histograms should provide a more accurate analysis, we will use the corrected histograms for further analysis.

\begin{figure*}
\begin{center}
\includegraphics[width=13cm]{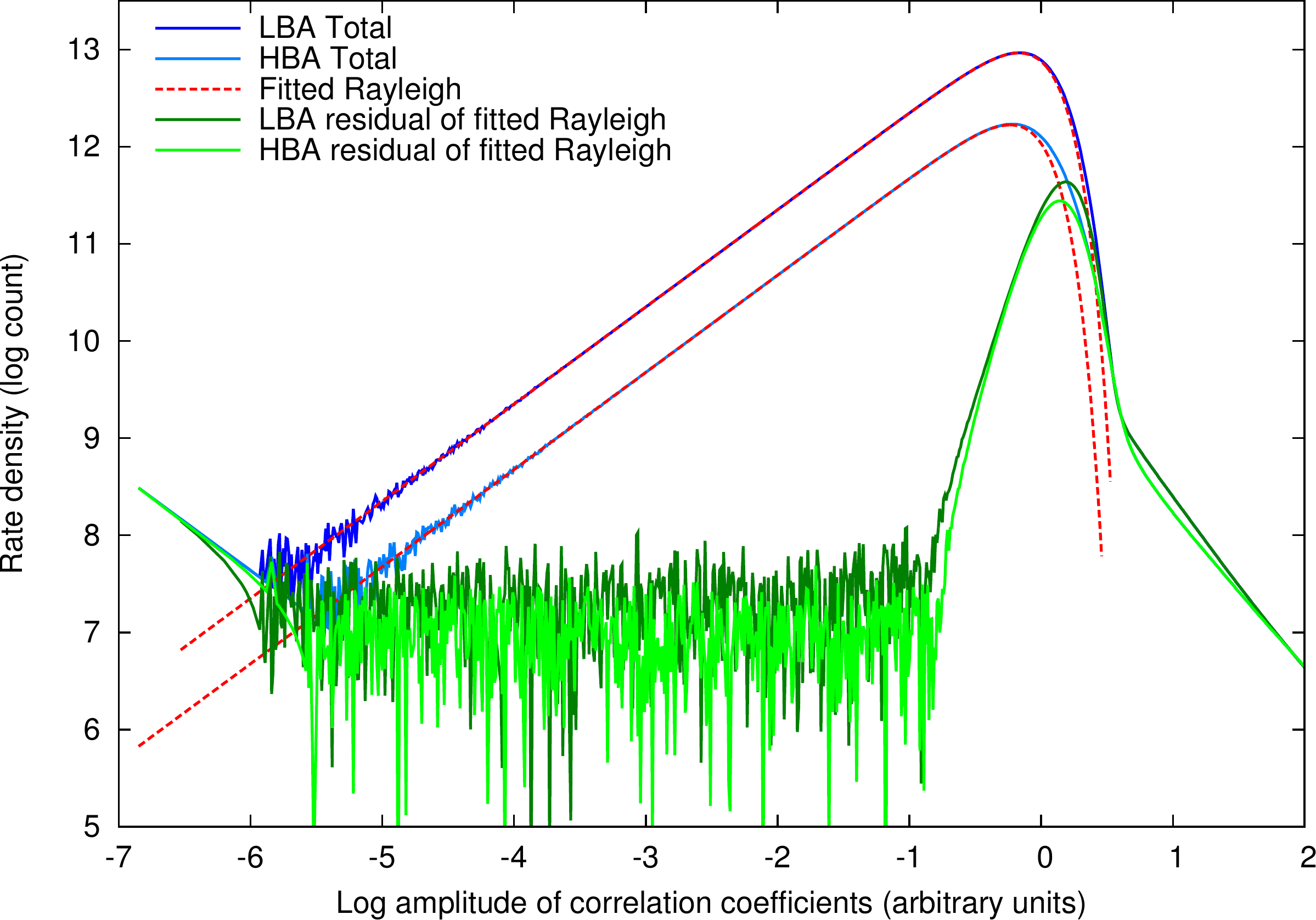}
\caption{Least-squares fits of Rayleigh distributions to the observed LBA and HBA histograms, after band-pass correction but without flux calibration.}
\label{fig:histograms-with-fits}
\end{center}
\end{figure*}

The Rayleigh parts of the distributions are plotted in Fig.~\ref{fig:histograms-with-fits}, along with a least-squares fit and its residuals. Both histograms follow the Rayleigh distribution for about five orders of magnitude, which is validated by the residuals that show only noise. It breaks down about one order of magnitude before the mode of the distributions. This is because of the multi-component nature of the distributions, as was described in \S\ref{sec:histogram-noise}.

\begin{figure*}
\begin{center}
\includegraphics[width=13cm]{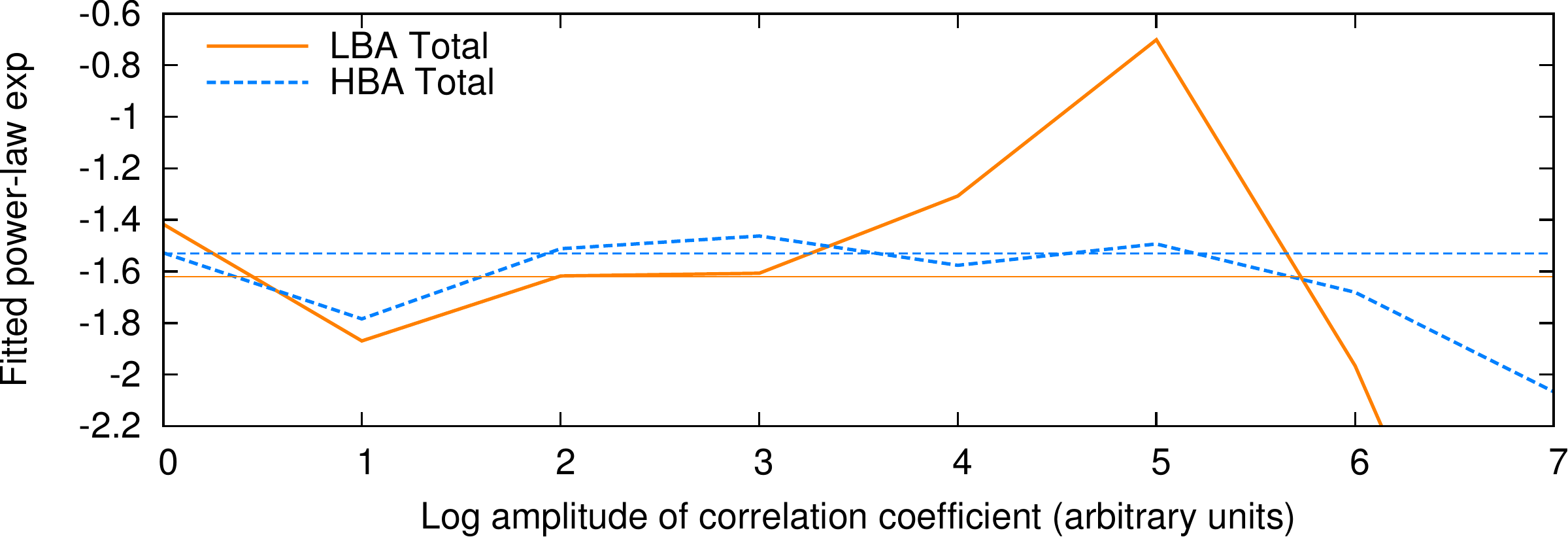}
\caption{The slope of the band-pass corrected log-log histogram as a function of the brightness. The horizontal lines indicate the fitted slope over the full (semi-) stable region. The horizontal axis is not calibrated.}
\label{fig:plot-slopes}
\end{center}
\end{figure*}

If we go back to Fig.~\ref{fig:histogram-passband-corrected}, we see that in the LBA the power law is stable for about three orders of magnitude, and one order more in the HBA. Fig.~\ref{fig:plot-slopes} shows the slope of the log-log plot as a function of amplitude, which was constructed by performing linear regression in a sliding window, with a window size of 1 decade. The HBA shows very little structure in the slope, but the LBA is less stable and shows some features in its power-law part. Linear regression on the power-law part of the log-log plot results in a slope of $-1.62$ for the LBA and $-1.53$ for the HBA. These and the other derived quantities have been summarized in Table~\ref{table:dist-data-quantities}. Although the HBA slope does not show any other significant features besides the Rayleigh and power-law curves, the LBA power law ends with a bulge around an amplitude of $10^6$. This bulge is caused by a very strong RFI source affecting lots of samples, and is a single outlier in the spatial distribution. We found this is caused by RFI observed for about an hour in the late afternoon in the lower LBA frequency regime, around $30$--$40$~MHz. Leaving this frequency range out flattens the bulge significantly, but does not completely eliminate it, because the source put the receivers in a non-linear state, causing leakage at lower intensity levels in the other sub-bands. Unlike linear regression, the fitting region of the Hill estimator is not limited at the high end. Consequently, because of the bulge, the Hill estimator evaluates for the LBA into a slope that is less steep, with a value of $-1.53$. For the HBA set, the Hill estimator is equal to the $-1.53$ value found by linear regression.

\begin{table*}
\centering
\begin{minipage}{12cm}
\caption{Estimated distribution quantities per data set. }\label{table:dist-data-quantities}
\begin{tabular}{@{}llrr@{}}
\textbf{Symbol} & \textbf{Description} & \textbf{LBA set}& \textbf{HBA set} \\
\hline
\hline
$N_\textrm{total}$ & Total number of samples in histogram & $8.0\times10^{11}$ & $5.4\times10^{11}$ \\
$\sigma$ & Rayleigh mode (assumed to be SEFD/$\sqrt{2\Delta t \Delta \nu}$, & $770$ Jy & $77$ Jy \\
 & where SEFD is the System Equivalent Flux Density) & & \\
\hline
\multicolumn{4}{l}{\textit{Estimators for power-law distribution parameters}} \\
\hline
$\alpha$ & Exponent of power law in RFI distribution      & $-1.62$ & $-1.53$ \\ 
$SE(\alpha)$ & Standard error of $\alpha$ & $2.8 \times 10^{-3}$ & $6.9 \times 10^{-4}$ \\
$\alpha_H$ & Hill estimator for power-law exponent & $-1.53$ & $-1.53$ \\
$SE(\alpha_H)$ & Standard error of $\alpha_H$ & $8.9 \times 10^{-6}$ & $1.0\times10^{-5}$ \\
$\epsilon_{\alpha}$ & Sampled estimate of standard deviation of $\alpha$ & $6.1 \times 10^{-2}$ & $1.2 \times 10^{-2}$ \\
$\beta$ & Scaling factor of power law with exponent $\alpha$ & $4.0 \times 10^{17}$ & $3.4\times 10^{15}$ \\
$\eta$ & Radiation fall-off speed for $\alpha$ ($\eta=2$ is free space) & 3.23 & 3.77 \\
\hline
\multicolumn{4}{l}{\textit{Limits}} \\
\hline
$S_L$ & Constraint on lower fall-off point of power law & $21$ mJy & $6.2$ mJy \\ 
$\tilde{S}_L$ & As $S_L$, but assuming $Ig/r^\eta\sim$ uniform & $47$ mJy & $14$ mJy \\ 
$S_d$ & Expected lowest apparent level of RFI detected & $26$ Jy & $5.7$ Jy\\ 
$E(S_R)$ & Apparent RFI flux density & $2,700$ Jy & $140$ Jy \\
$E(S_\textrm{leak})$& Residual apparent RFI flux density after excision & $484$--$496$ mJy & $167$--$171$ mJy \\
 & Same as above, but by assuming 10\% occupancy & $384$ mJy & $120$ mJy \\ 
REFD& RFI equivalent flux density & $18.9$--$19.3$ Jy & $6.5$--$6.7$ Jy \\
\hline
\multicolumn{4}{l}{\textit{Average station temperatures}} \\
\hline
$T_\textrm{sys}$ & System temperature (in clean bands)  & 5,000 K & 640 K \\
$T_\textrm{R}$ & RFI Temperature & 17,000 K & 1,200 K \\
$T_\textrm{leak}$ & Temperature of undetected RFI & 3.2 K & 1.4 K \\
\hline
\hline
\end{tabular}
\end{minipage}
\end{table*}

On the assumption that the histogram is zero below amplitude $S_L$, we find that $S_L=21$ mJy for the LBA and $S_L=6.2$ mJy for the HBA (see Table~\ref{table:dist-data-quantities}). If instead it is assumed that the histogram has a uniform distribution below some amplitude $\tilde{S_L}$, we find that the amplitude at which the power-law distribution breaks down is approximately a factor two higher. The two different assumptions on how the power-law distribution breaks down have a small effect on $E(S_\textrm{leak})$, the expected value of the leaked RFI. By using $\tilde{S}_L$ instead of $S_L$, it is a few percent lower. By assuming a 100\% RFI occupancy, we find that the expected value of leaked RFI is $484$--$496$ mJy for the LBA and $167$--$171$ mJy for the HBA. By assuming $10$\% occupancy, the value for $E(S_\textrm{leak})$ is about $25$\% reduced. The RFI occupancy only starts to have a significant effect on $E(S_\textrm{leak})$ if it is well below $10$\%. 

\section{Conclusions and discussion} \label{sec:dist-discussion}
We have analysed the histogram of visibility amplitudes of LOFAR observations and found that, within a significant range of the histogram, the contribution of RFI sources follows a power-law distribution. The found power-law exponents of $-1.62$ and $-1.53$ for the 30--78~MHz LBA and 115--163~MHz HBA observations respectively, can be explained by a uniform spatial distribution of RFI sources, affected by propagation described surprisingly well by Hata's electromagnetic propagation model. Taken at face value these exponents imply in Hata's model that the average transmitting heights for sources affecting the LBA and HBA are $79$ and $13$~m respectively. There are no 79~m high transmitters nearby LOFAR stations in the LBA frequency range. Additionally, Hata's model only goes down to 150~MHz, and it is possible that the electromagnetic fall-off due to propagation will be different for lower frequencies. Intervals for the exponents with representative $3\sigma$ boundaries are $[-1.80;-1.44]$ for the LBA and $[-1.57;-1.49]$ for the HBA, giving average transmitter heights of $[0.6; 800]$ and $[3.1; 23]$~m for the LBA and HBA respectively. Therefore, the LBA measurements are clearly not accurate enough to be conclusive. Moreover, because the power-law distribution analyses involve many assumptions, it is uncertain whether the analyses are sufficiently accurate for making these detailed conclusions.

On the assumption that the power-law distribution for RFI sources will continue down into the noise, we have constructed a full parametrization of the RFI apparent flux distribution. By assuming that all samples contain some contribution of RFI, we find that the average flux density of RFI after excision by automated flagging is $484$--$496$~mJy for the LBA and $167$--$171$~mJy for the HBA. These values should be compared to the noise in individual samples of $770$~Jy (LBA) and $77$~Jy (HBA) (see Table~\ref{table:dist-data-quantities}), and are upper limits for what can be expected. If in fact not all samples are affected by RFI, the leaked RFI flux will be smaller, and will of course be zero in the extreme case that the detector has found and removed all RFI.

The apparent RFI flux densities can be converted to a RFI station temperature that excludes the system noise and sky noise components. If we use a station efficiency factor $\eta_\textrm{st}=1$ and effective areas LBA $A_\textrm{eff}=398$ and HBA $A_\textrm{eff}=512$ with again 100\% RFI occupancy, our models lead to RFI temperatures of 17,000~K and 1,200~K for respectively the LBA and the HBA. These are relatively high compared to for example the survey by \citet{lf-interference-temperature-rogers}, who report that on two different sites, 20\% and 27\% of the spectrum has a temperature above 450~K in the range of 50--1500~MHz. However, our post-detection RFI station temperatures, which arise from the residual apparent RFI flux density estimates, are 3.2~K and 1.4~K for the LBA and HBA respectively. Due to LOFAR's high resolution and accurate flagging strategy, this is achieved by flagging a relatively small data percentage of 1.8 (LBA) and 3.2\% (HBA).

In projects such as the EoR detection experiment with LOFAR, a simulation pipeline is used to create a realistic estimate of the signal that can be expected. Currently, these simulations do not include the effects of RFI. With the construction of empirical models for the RFI source distributions, we are one step closer to including these effects in the simulation. Using Eq.~\eqref{eq:drawing-rfi-with-propagation}, one can sample a realistic strength of a single RFI source, add the feature to the data and run the AOFlagger. What is still needed for accurate simulation, is to obtain a likely distribution for the duration that one such source affects the data. For example, it is neither realistic that all RFI sources are continuously transmitting nor that they affect only one sample. The RFI detector is highly depending on the morphology of the feature in the time-frequency domain. Finally, the coherency properties of the RFI might be even more important to simulate correctly, but these have been not been explored. However, these have large implications for observations with high sensitivity. This will be discussed in the next section. 

The derived values for the average lower level of detected RFI, $S_d$, show that the AOFlagger has detected a large part of the RFI that is well below the sample noise. In both sets, $S_d$ is more than one order of magnitude below the Rayleigh mode. This can be explained with two of the algorithms it implements. The first one is the {\tt SumThreshold} method \citep{post-correlation-rfi-classification}, that thresholds on combinations of samples, and is thus able to detect RFI that is weaker than the sample noise. The second one is the scale-invariant rank (SIR) operator \citep{scale-invariant-rank-operator}. This operator is not dependent on the sample amplitude, but flags based on morphology.

\subsection{Implications for very long integrations}
Faint RFI could impose a fundamental limit on the attainable noise limit of long integrations. We will analyse the situation for the LOFAR EoR project. This project will use the LOFAR high-band antennas to collect on the order of 50--100 night-time observations of 6 h for a few target fields. The final resolution required for signal extraction will be about $1$~MHz. The project will use about $60$~stations, each of which provides two polarized feeds. This will bring the noise level in a single 6~h observation in 1~MHz bandwidth to
\begin{equation}
 \sigma_\textrm{eor-night} = \textrm{SEFD} \left(2\Delta t \Delta \nu N_\textrm{feed} N_\textrm{interferometers} \right)^{-\frac{1}{2}} \approx 250\hspace{1mm} \mu\textrm{Jy},
\end{equation}
where $N_\textrm{feed}=2$ is the number of feeds per antenna and $N_\textrm{interferometers}=\frac{1}{2}60\times59$ is the number of interferometers. Therefore, after 100 nights the thermal noise level will be $25\hspace{1mm}\mu\textrm{Jy}$.

Because some RFI sources might be stationary, the signals from these sources will add consistently over time, meaning that the geometrical phase will be the same every day. Therefore, the amount that time integration can decrease the flux density of RFI might be limited. On the other hand, many RFI signals observed in the LOFAR bands have a limited bandwidth, and the majority of the detected RFI sources affect only one or a few LOFAR channels of 0.76 kHz. Therefore, frequency averaging will lower the flux density of the RFI signal. If the frequency range contains only one stationary RFI source, the strength of this source will go down linearly with the total bandwidth. If we assume that all channels are affected by RFI sources and all these sources transmit in approximately one channel, then the noise addition that is produced by RFI will go down with the square root of the number of averaged channels. This is a consequence of the random phase that different RFI sources have.

In summary, some class of stationary RFI sources are expected to add consistently over time, polarization and interferometer, but not over frequency. Therefore, in this case the noise level at which RFI leakage approximately becomes relevant is given by
\begin{equation} \label{eq:rfi-noise-level}
 \sigma_\textrm{RFI} = \frac{\textrm{REFD}}{\sqrt{2 \Delta \nu}},
\end{equation}
where $\textrm{REFD}$ is the RFI equivalent flux density at $1$ Hz and $1$ s resolution for a single station, in analogue to how the SEFD is defined. This only holds when the observational integrated bandwidth $\Delta \nu$ is substantially higher than the average bandwidth of a single RFI source. The empirically found upper limits in this work are $\textrm{REFD}_\textrm{LBA}=18.9$--$19.3$~Jy and $\textrm{REFD}_\textrm{HBA}=6.5$--$6.7$~Jy (see Table~\ref{table:dist-data-quantities}).

For the EoR project with $1$~MHz resolution, Eq.~\ref{eq:rfi-noise-level} results in $\sigma_\textrm{RFI}\approx 4.7$ mJy. However, the first EoR results of observations of one day have approximately reached the thermal noise of about $1.7$ mJy per 0.2~MHz sub-band \citep{ncp-eor-yatawatta}, and the resulting images show no signs of RFI. This implies that either the upper limit is far from the actual RFI situation, or Eq.~\ref{eq:rfi-noise-level} is not applicable to most of the RFI that is observed with LOFAR. In the following section we will discuss effects that could cause a reduced contribution of RFI.

\subsection{Interference-reducing effects} \label{sec:coherence-reduction}
When integrating data, it is likely that the actual noise limit from low-level RFI will be significantly lower than the given upper limit, which was determined at highest LOFAR resolution. There are several reasons for this: Many RFI sources have a variable geometric phase, because they move or because their path of propagation changes; many RFI sources will be seen by only a few stations or are not constant over time; for the shortest baselines at 150~MHz, the far field starts around 1~km, and some RFI sources will therefore be in the near field; and finally, a large number of stationary RFI sources in a uniform spatial distribution will interfere both constructively and destructively with each other. These arguments are valid only for interferometric arrays. Global EoR experiments that use a single antenna will not benefit from these effects, and will still be limited by low-level RFI.

Fringe stopping interferometers can partly average out RFI sources. Nevertheless, stationary RFI that is averaged out by fringe stopping will leave artefacts in the field centre \citep{post-correlation-filtering}. This is not relevant when observing the North Celestial Pole --- which is one of the LOFAR EoR fields --- because no fringe stopping is applied when observing the NCP. Imaging of the data will localize the contribution from stationary RFI near the NCP. If RFI artefacts would show in the image of the NCP field, they can easily be detected and possibly be removed, or processing could ignore data near the pole. Because of these arguments, it is a risk to use the NCP as one of the EoR target fields. At the same time, this field is useful for analysing the RFI coherency properties. Preliminary analysis of EoR NCP observations of a single night have almost reached the thermal noise, but do not show leaked RFI at the pole \citep[\S4.3]{ncp-eor-yatawatta}.

Because we have assumed 100\% of the spectrum is occupied by RFI, our given RFI constraints could be too pessimistic. If only 10\% of the samples are affected by RFI, the expected value of the leaked RFI level decreases by about $25$\%, and if the detected 2.68\% true-positives contain all RFI, there is no leaked RFI at all. With current data, one can only speculate how much of the electromagnetic spectrum is truly occupied.

Finally, future RFI excision strategies can further enhance detection accuracy. Once data from a large number of nights are collected, it will be possible to detect and excise RFI more accurately. With the current strategy, it is likely that the LOFAR EoR project will encounter some RFI on some frequencies when averaging lots of observing nights, although this still remains to be seen. To mitigate this leaked RFI, the detection can be executed at higher signal-to-noise levels. The current results indicate that a lot of RFI does not add up consistently, and the situation is promising. Considering the current RFI results, and the availability of further mitigation steps, we conclude that RFI will likely not be problematic for the detection of the Epoch of Reionisation with LOFAR.

\section*{Acknowledgments}
LOFAR, the Low-Frequency Array designed and constructed by ASTRON, has facilities in several countries, that are owned by various parties (each with their own funding sources), and that are collectively operated by the International LOFAR Telescope (ILT) foundation under a joint scientific policy. Parts of this research were conducted by the Australian Research Council Centre of Excellence for All-sky Astrophysics (CAASTRO), through project number CE110001020. C. Ferrari and G. Macario acknowledge financial support by the ``Agence Nationale de la Recherche'' through grant ANR-09-JCJC-0001-01.

\DeclareRobustCommand{\TUSSEN}[3]{#3}

\bibliographystyle{mn2e}
\bibliography{references}

\label{lastpage}

\end{document}